# Deep learning search for gravitational waves from compact binary coalescence

Lorenzo Mobilia,[1,2] Tito Dal Canton,[3] and Gianluca Maria Guidi[1,2]

[1]*Università degli Studi di Urbino Carlo Bo, 61029 Urbino, Italy*
[2]*INFN, Sezione di Firenze, 50019 Sesto Fiorentino, Firenze, Italy*
[3]*Université Paris-Saclay, CNRS/IN2P3, IJCLab, 91405 Orsay, France*

(Received 14 March 2026; revised 3 September 2026; accepted 1 October 2026; published 21 October 2026)

Gravitational-wave searches rely on a combination of methods, including matched filtering, coherent analyses, and more recent machine-learning–based pipelines. For compact binary coalescences—where signals originate from the relativistic dynamics of compact objects—matched filtering remains a central element, but its computational cost will increase substantially with the data volumes and parameter-space coverage required by next-generation interferometers such as the Einstein Telescope. Developing complementary strategies that reduce computational load while preserving detection performance is therefore essential. We investigate a hybrid approach that combines matched-filtering concepts with convolutional neural networks, enabling efficient signal searches without relying on the usual $\chi^2$ rejection test. Using simulated datasets that include injected signals in Gaussian noise, transient noise, and physical effects not represented in the template bank—such as eccentricity, precession and higher-order modes—we show that the method achieves a detection efficiency comparable to or better than a standard matched-filtering search while offering a more resource-efficient pipeline in all the cases where the physics included in the signals were not represented in the template bank or when glitches were affecting the data. Conversely, the analysis also reveals some loss of detection capability for standard binary neutron star and binary black hole searches in pure Gaussian noise. These results indicate that deep-learning–assisted searches can support sustainable gravitational-wave data analysis in future detector eras.

## I. INTRODUCTION

Gravitational Wave (GW) astronomy began with the first detection, GW150914 [1] produced by a binary black hole (BBH) system. Since then, the field has grown considerably, with more than 200 GW detections [2–6] up to the first part of the fourth data acquisition campaign (O4a), and many more expected from the final analysis of the most recent run, O4. The scientific output of the field has been correspondingly rich. Events such as GW170817 [7], the first binary neutron star (BNS) signal ever detected, and GW190521 [8], a signal from an intermediate-mass black hole, have opened the door to multimessenger astronomy with GWs and deepened our understanding of the possible astrophysical origins of massive black holes [9] that are incompatible with classical stellar formation scenarios.

These detections have been made possible thanks to the extensive effort of the LIGO-Virgo-KAGRA (LVK) Collaboration in maintaining and enhancing the sensitivity of the interferometers: LIGO [10] in the U.S., Virgo [11] in Italy, and KAGRA in Japan [12]. Looking ahead, another interferometer is planned to be built in India [13], and several observatories targeting different frequency bands are under development. These include the space-based LISA mission [14,15], the Einstein Telescope [16] in Europe, and the Cosmic Explorer [17] in the U.S. Alongside the technical and technological advancements driven by the hardware side, a significant effort in detecting GW signals is carried out by data analysis techniques. The vast majority of detected events are compact binary coalescences (CBCs), including the aforementioned BBH and BNS, as well as mixed systems such as neutron star–black hole (NSBH) binaries. These signals are identified by constructing a bank of theoretical waveforms, the template bank, derived from the solutions of the Einstein equations, and applying matched filtering to the data recorded by the detectors. Several pipelines implement this approach, including GstLAL [18–21], PyCBC [22], MBTA [23], and SPIIR [24]. In addition to these modeled searches, unmodeled searches are also conducted by identifying correlated excesses of energy across detectors, as done in the coherentWaveBurst pipeline [25,26].

In this context, several machine-learning approaches to gravitational-wave detection have been investigated. These span end-to-end architectures operating directly on the whitened strain [27–29], as well as hybrid schemes in which a network is applied downstream of a matched-filtering stage. The first machine learning gravitational-wave search mock data challenge [30] provided a common benchmark for these methods, and found that the best machine-learning entries reach up to 95% of the sensitive distance of production matched-filtering analyses in simulated Gaussian noise, with a larger gap on real detector data. Since then, the field has moved toward demonstrations on real observing-run data: [31,32] present a deep-learning pipeline for binary neutron star signals and binary black holes, and [33] proposes a hybrid algorithm in which a network is trained on a map of matched-filter SNR time series, an approach closely related to the one adopted here. The present work differs from these studies, as well as from standard matched-filtering searches, in three respects. First, the input to the network is built from a deliberately reduced, one-dimensional template bank spanning chirp mass alone, rather than from a bank covering the full intrinsic parameter space; the computational saving this affords is the primary motivation for the method. Second, whereas production matched-filtering pipelines rely on a $\chi^2$ signal-consistency test to suppress glitch-induced triggers [34,35], at no stage do we compute such a test; we instead rely on the network to learn the distinction between signal-like and glitchlike structures directly from the map. Third, the emphasis of our simulation campaigns is on signals whose physics is deliberately absent from the template bank—spin, precession, higher-order modes, eccentricity, and overlapping events—which is the regime where a consistency test is expected to be least reliable. We stress that the results presented here are obtained entirely on simulated Gaussian noise with synthetic transients, and a demonstration on real detector data remains necessary before the method can be assessed as a search pipeline. The paper is structured as follows: In Sec. II we describe the matched-filtering technique used to detect GW signals, along with the usual statistical rejection test applied to the candidates to remove spurious transient noise. In Sec. III, we define the time-template signal-to-noise-ratio (SNR) map (TT-SNR map) as the observable obtained by stacking the SNR time series from the matched-filtering output of a template bank. A description of the template bank, along with the construction procedure, is also provided. In Sec. IV we briefly introduce the fundamental elements of the residual network algorithm implemented in this study. In Sec. V we present the first simulation campaign, which considers only Gaussian noise and binary-neutron star system waveforms. In Sec. VI we enlarge the parameter space to include binary-black holes systems and introduce transient noise artifacts (also known as glitches) in the data to test the robustness of the network against non-Gaussian noise. We also compare the network output with the standard detection significance. In Sec. VII we include spinning signals, to assess the network's ability to recognize more physically complex waveforms. In Sec. VIII we expand the analysis by including signals with precession, eccentricity, and higher-order modes. In addition, extreme-spin scenarios and superimposed signals are considered. Finally, in Sec. IX we summarize the conclusions and discuss possible extensions and future applications of this study.

## II. MATCHED-FILTERING SEARCHES

Detecting GW signals is challenging. The signals we aim to detect are usually orders of magnitude weaker than the noise level in the interferometer's strain data. For this reason, we rely on the matched-filtering technique, which allows us to probe deep into the noise floor by providing an observable measure of the loudness of a signal buried in noise [36]. Let us define the interferometer's data strain time series $s(t)$. Matched-filtering involves correlating this quantity with a set of precomputed templates $h(t)$—obtained from solving the Einstein field equations for binary systems in general relativity using mathematical models known as approximants—and weighting the correlation by the detector's noise power spectral density $S_n(f)$. This procedure yields a SNR time series, $\rho(t)$, which quantifies the amplitude of a candidate GW signal relative to the noise floor. $\rho(t)$ is defined as

$$\rho(t) = \frac{(s|h)(t)}{\sqrt{(h|h)(t=0)}}, \tag{1}$$

where $(s|h)(t)$ is the real part of the noise-weighted inner product

$$(s|h)(t) = 4\mathcal{R} \int_0^\infty \frac{\tilde{s}(f)\tilde{h}^*(f)}{S_n(f)} e^{2i\pi f t} df. \tag{2}$$

Here, $\tilde{s}(f)$ and $\tilde{h}(f)$ denote the Fourier transforms of the data strain $s(t)$ and the template signal $h(t)$. This expression correlates the data with the template at a fixed reference phase. Since the coalescence phase of the signal is unknown, it is common to compute instead the complex matched-filter

$$z(t) = 4 \int_0^\infty \frac{\tilde{s}\tilde{h}^*}{S_n(f)} e^{2i\pi f t} df, \tag{3}$$

whose real and imaginary parts correspond to correlations with templates 90° out of phase. The phase-independent detection statistic is then obtained by taking the modulus of Eq. (3) and normalizing it by the template norm $\sqrt{(h|h)(t=0)}$,

$$\rho(t) = \frac{|z(t)|}{\sqrt{(h|h)(t=0)}}. \quad (4)$$

The classical CBC search consists of computing Eq. (4) for a set of templates organized into a template bank. A maximum of $\rho(t)$ is then selected within a given time window and across the entire template bank; if this value exceeds a predefined threshold, the event is flagged as a candidate detection. In purely Gaussian noise, and assuming that all signal parameters except the amplitude are known, matched filtering is proven to be the optimal detection statistic. In practice, since these parameters are not known *a priori*, the resulting filter is only approximately optimal [36].

The matched-filtering based searches require template banks that adequately cover the parameter space of interest. In fact, Eq. (2) can be read as a scalar product, that tells us how similar the template $h$ and the signal $s$ are. It follows that if they differ substantially, the recovered SNR is significantly reduced. Conversely, the largest SNR is obtained when the template coincides with the signal contained in the data. For a signal $h_s(t)$, this defines the optimal SNR

$$\rho_{\rm opt} = \sqrt{(h_s|h_s)(t=0)}, \quad (5)$$

i.e., the SNR that would be recovered by filtering noise-free data with a perfectly matching template. Equation (5) is formally identical to the normalization factor of $z(t)$ in Eq. (4), the only difference being that it is evaluated for the signal actually present in the data rather than for a template of the bank. In this work, $\rho_{\rm opt}$ is computed for each injected waveform with the same power spectral density (PSD) used to color the noise, and it is used to quantify the intrinsic loudness of the injections independently of the template bank. The choice of the template bank is therefore a fundamental step in CBC searches. Constructing such a bank—especially in high-dimensional parameter space involving, for example, masses and spins—can be challenging. Several algorithms have been developed for this purpose, including purely geometric approaches such as |`pycbc_geom_aligned_bank`| [37,38], stochastic-geometric hybrid methods [39], and even fully stochastic approaches [40].

In practice, however, searches must contend with noise that is not purely Gaussian. If transient noise (glitches) affects the data stream, the matched-filtering can produce SNR values similar to those produced by genuine signals, making the detection of real astrophysical sources challenging [41]. For this reason, LVK search pipelines have developed techniques to face this challenge. Some pipelines (such as MBTA, GstLAL and SPIIR), following [18], exploit an observable known as the autocorrelation function to penalize glitchlike trigger behavior by reducing the statistical significance of the corresponding events. Here, to compare the results of the network with a statistics that takes into account this issue, we adopt a technique that suppresses glitch-induced SNR by applying a statistical discriminator: the $\chi^2$ test (or its reduced form $\chi^2_{\rm r}$), originally proposed in [34] and defined as

$$\chi^2_{\rm r} = \frac{\chi^2}{(2N_b) - 2}, \quad (6)$$

where $N_b$ is the number of frequency sub-bands used in the test. In this study, $N_b$ has been set to 16. Even though this is not the optimal choice—the number of sub-bands is a function of the template [42]—we chose to fix it for practical reasons linked to the computational cost. The idea behind this is to build a statistic that indicates if the filter output is consistent with the expected signal in the frequency interval of interest. The $\chi^2$ time-frequency discriminator has proven to be an effective method in reducing the impact of transient noise on $\rho(t)$ by implementing the so-called reweighted-SNR $\rho_{\rm rw}$, following the procedure introduced in [35]:

$$\rho(t)_{\rm rw} = \frac{\rho(t)}{\left[\frac{(1+\chi^2_{\rm r})^2}{2}\right]^{\frac{1}{6}}}. \quad (7)$$

This new quantity is then used as an effective statistic to discriminate genuine astrophysical signals from spurious noise. By applying several possible cuts, first of all selecting all the triggers above a minimal threshold $\rho_{\rm thr}$, a distribution of signal and noise is obtained. This statistic is then commonly used to define the false alarm rate, that is the rate at which noise produces a candidate ranked equal to or above the one we are considering.

We remark that this technique is not the only one proposed as SNR modeling for a glitch-rejection test. Other possibilities explored are the auto-$\chi^2$ test in [18], where the SNR time series of the data is compared to the time series expected from a real signal, and the bank veto in [43]. In addition, a possible unified approach to the $\chi^2$ test is explored in [44].

The use of the $\chi^2$ test is essential for efficiently rejecting spurious noise. On the other hand, if the template we are considering for this time-frequency discriminator does not perfectly match the signal present in the data, the resulting $\rho(t)_{\rm rw}$ will also be impacted, resulting in a down ranking that can lead to a suboptimal detection, or even missing the signal if the statistic is pushed below the threshold. This means that both a bank adequately covering the parameter space and a careful calibration of the $\chi^2$ test are required. In this study, we explore the possibility of detecting GW signals using a reduced template bank and avoiding explicit rejection-test discriminators by adopting a deep-learning approach that autonomously provides a rejection criterion through learning the characteristic features of signal and noise present in the data.

## III. THE TIME-TEMPLATE SNR MAP

Classical searches involve hundreds of thousands or even millions of templates [45]; together with the calibration of the $\chi^2$ test, this makes the generation of candidate triggers computationally demanding. In this work, we build a small template bank and apply the matched-filtering process that produces an SNR time series $\rho(t)$ for each template in the template bank. Hence we investigate the possibility of identifying a candidate event by simultaneously analyzing the $\rho(t)$ values across the entire template bank within a time window of a few seconds, avoiding the computation of the $\chi^2$.

Matched filtering correlates each template with the stretch of data being analyzed. We expect that the SNR time series will have a different profile for each template we are considering, encapsulating both the physical parameters of the signal and the merging time of the signal. To illustrate this, let us consider a signal characterized by a reference chirp mass $M_c^{\rm ref}$, where the chirp mass is defined as

$$M_c = \frac{(m_1 \cdot m_2)^{\frac{3}{5}}}{(m_1 + m_2)^{\frac{1}{5}}}, \tag{8}$$

where $m_1$ and $m_2$ are the component masses, measured in solar masses $M_\odot$. Systems with chirp masses larger or smaller than $M_c^{\rm ref}$ evolve through the inspiral at different rates and therefore reach merger at earlier or later times, respectively [46]. Consequently, the matched-filter SNR time series is expected to exhibit systematic variations in its temporal profile as a function of the template chirp mass. By simultaneously analyzing the $\rho(t)$ of the whole bank, we are able to infer the presence of a signal in the data from the overall morphology of that distribution (see the Appendix for a pictorial representation of this phenomenon). In addition, we expect glitches to have a completely different structure in this correlation, avoiding, in principle, the use of the $\chi^2$ test.

To do so, we construct a TT-SNR map, obtained by stacking the $\rho(t)$ time series from all templates in the bank. If a signal is present in the data, then the TT-SNR map exhibits characteristic structures that can be recognized by machine-learning algorithms, such as a convolutional neural network (CNN) [47]. We expect that even if templates are not optimal—i.e., they can significantly differ from the signal—this would not significantly impact the TT-SNR map, and the CNN would be able to learn such behavior.

To this end, we construct a one-dimensional template bank targeting equal-mass binary systems, including both BNS and BBH. The parameter space is defined in terms of chirp mass $M_c$, so that we are able to detect signals produced by several systems. A key metric characterizing the template bank is the fitting factor [48]. The GW parameter space is continuous, whereas the template bank must be constructed from a finite and discrete set of points. This discretization introduces a mismatch, as the templates in the bank typically do not match the signal exactly. The resulting loss in SNR is quantified by the fitting factor, which measures the fractional loss in SNR due to the bank's sparsity. In this work we choose a maximum SNR loss of 3%.

The template placement algorithm determines the optimal chirp mass spacing between adjacent templates by ensuring that the match between them is equal to the minimum required match, here set to 0.97 with a lower frequency cutoff of 30 Hz. It considers symmetric binary systems ($m_1 = m_2$) and iteratively computes the spacing in chirp mass between a reference template, and a candidate template. The masses span from $m_{1,2} \in [1.0 - 150] M_\odot$; consequently the chirp mass is $M_c \in [0.87\text{–}130.6] M_\odot$. The match is computed using the detector PSD, here modeled by `aLIGOZeroDetHighPower` [49] that is an ideal design sensitivity of Advanced LIGO. To compute the match between adjacent templates, the waveforms are generated using the approximant `SEOBNRv4_ROM` [50], i.e., the same model later used to filter the data. Applying this procedure results in a one-dimensional template bank consisting of 5,448 templates.

In this study, to construct the TT-SNR maps, we inject a GW signal into 80-second-long data segments using the `SEOBNRv4_opt` approximant [51–53] generated from 27 Hz in Gaussian noise colored by `aLIGOZeroDetHighPower` power spectral density. The data are then matched filtered against the one-dimensional template bank generated with the `SEOBNRv4_ROM` [50,52] model, using a frequency range $\Delta f \in [30\text{–}1024]$ Hz with the same PSD used to color the Gaussian noise and construct the bank. The software used for injecting signals and computing the matched filtering is PyCBC [54]. We can then build the TT-SNR map in this way: Once the $\rho(t)$ time series is computed for each template we extract a 4-second time window around the true merger time of the injected signal with a sampling rate of 2048 Hz. The window is not exactly centered on the true merger time: its center is displaced by a random time jitter, drawn from a uniform distribution within $\pm 0.2$ s, so that the merger does not always fall on the same pixel column of the image and the network cannot exploit a fixed merger position as a spurious classification feature. This produces an $M \times N$ matrix, where M is the number of templates in the bank while N is the number of time samples in the window, i.e., $4 \times 2048$. In this way each column of the matrix corresponds to a different time sample of the SNR time series, and each row corresponds to a different template. Since CNN training is sensitive to input image dimensions, we compress the TT-SNR map into a $512 \times 256$ grayscale image. The compression is performed with bilinear sampling using the PIL Python module. This results in a lossy operation, and sets a floor on the amount of structure the CNN can exploit: Inevitably, features of the SNR distributions are smeared away when those are narrower than the effective pixel size before they reach the first convolutional layer. The choice of $512 \times 256$ is dictated by the memory footprint of the training stage on the CPU resources available for this study. We return to the consequences of this compression in Sec. V. To standardize the representation, the SNR is mapped to the unit interval by applying a logarithmic normalization procedure. By selecting as normalization bounds the values $[b_{\rm min} = 0.1$–$b_{\rm max} = 100]$ we then compute $\bar{\rho}(t) = \frac{\log(\rho(t)) - \log(b_{\rm min})}{\log(b_{\rm max}) - \log(b_{\rm min})}$ which maps $\bar{\rho}(t)$ into $[0-1]$. In the resulting grayscale image, high-SNR values appear as brighter pixels, while low-SNR values correspond to darker shades of gray. Examples of these TT-SNR maps are shown in Fig. 1.

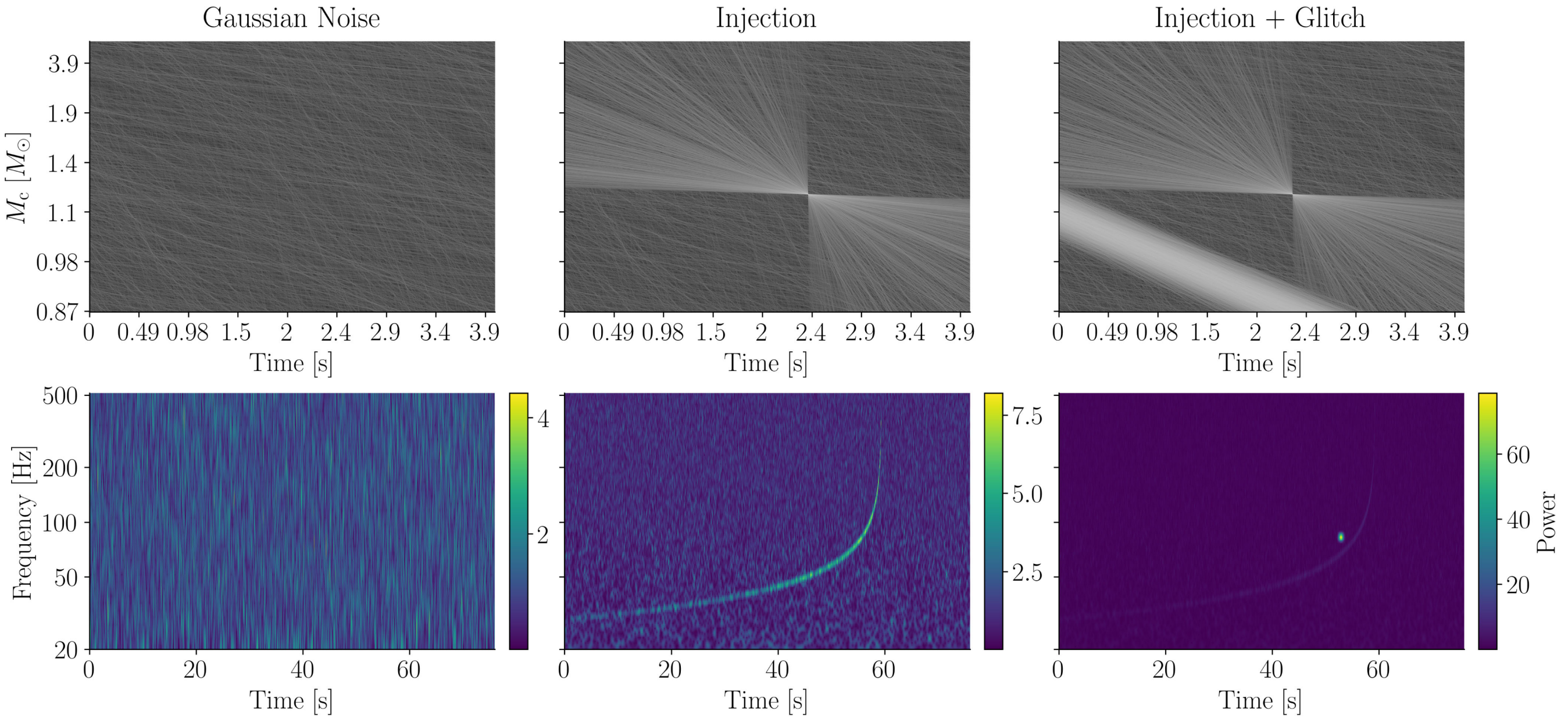

FIG. 1. Examples of the TT-SNR map (top) obtained by piling up the $\rho(t)$ after the normalization procedure along with its spectrogram representation (bottom). The left panel shows Gaussian noise only, the center panel an injected signal, and the right panel a signal with a superimposed glitch. The injected signal consists of a BNS system of 1.4–1.4$M_\odot$ at 50 Mpc ($\rho_{\rm opt} = 65.1$), while the glitch is a sine-Gaussian signal generated with parameters $f_0 = 82.74$, $Q = 232.11$, Amplitude $A = 8.82 \times 10^{-22}$, $\phi_0 = 4.01$ injected at 8.08 s before the merger time of the BNS signal.

The core idea is that the CNN can learn the feature-rich representation provided by the TT-SNR map to distinguish between noise and injected signals. This reframes the classical GW triggering problem as an image recognition task, where CNNs are known to perform well. The images, labeled as "noise" or "signal" according to whether a signal is present in the data, are used to train a CNN to classify each input into the two classes by assigning a score $r \in [0, 1]$ (or CNN output), where values near 0 correspond to noiselike classifications, and values close to 1 indicate the presence of a signal. In this work we considered different datasets with varying noise properties and signal parameters, resulting in four different simulation studies. In general, the dataset is split such that 70% is used for training, while the remaining 30% is equally divided between validation and test sets.

## IV. RESIDUAL NETWORK ALGORITHM

The idea of applying automated learning algorithms inspired by the human brain has its roots in Rosenblatt's perceptron model [55]. The original idea was to mimic the chemical activity of a biological neuron using a simple function that becomes activated in response to a given input. Modern convolutional neural networks are built upon this fundamental concept. Their typical architecture consists of multiple layers of interconnected neurons. The learning process is in general based on minimizing a loss function that quantifies the model's performance on a given training dataset. By adjusting the model's internal parameters—using the back-propagation algorithm to minimize the loss [56]—the network can learn complex relationships within the data. This allows the model to generate predictions that are consistent with the expected outputs for a given task. A convolutional network is an algorithm based on the convolutional layer that applies convolution to the input data [57]. This process requires the application of filters known as kernels that are responsible for learning complex patterns from images.

In this work, we adopt a specific architecture known as residual network (ResNet) [58]. This model is explicitly designed for image recognition tasks and has been optimized for such purposes through the residual block. ResNet introduces the concept of residual blocks, whose key idea is to let the network learn residual corrections rather than complete transformations. The key feature of the residual block is that the input passes through several convolutional layers where it is processed, and the result is then added back to the original input by a shortcut connection and a summation block. These identity shortcuts stabilize the training process and improve convergence, by mitigating the vanishing-gradient problem, making ResNet a state-of-the-art architecture in deep learning for image recognition. In our case, this flexible model, which we refer to as EasyResNet, is trained specifically to distinguish TT-SNR maps generated from noise from those containing a signal. The structure of the network used in this work is depicted in Fig. 2.

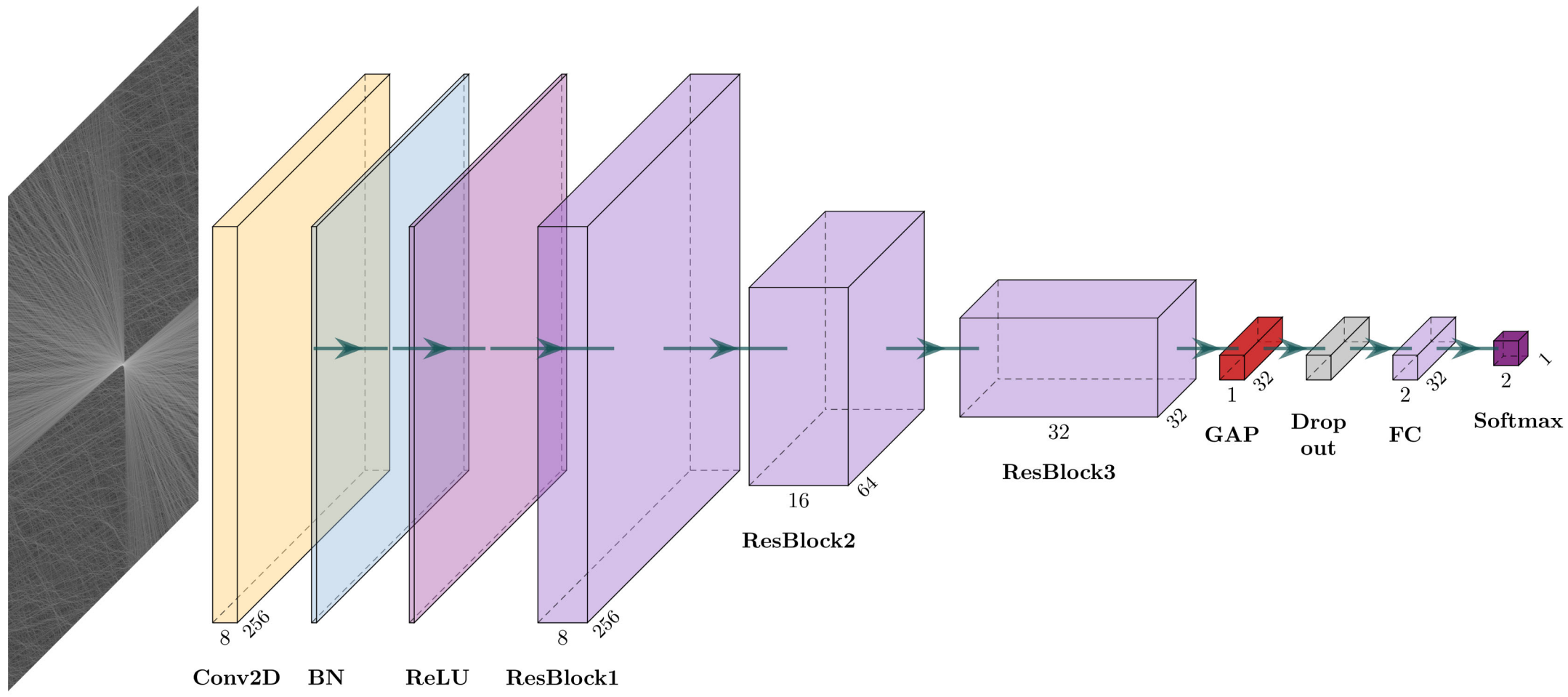

FIG. 2. EasyResNet architecture. A convolutional stem (Conv2D-BN-ReLU) processes the input and feeds three residual blocks (ResBlock1-3). The final feature maps are reduced by global average pooling (GAP) to a feature vector, regularized with dropout, and passed to a fully connected (FC) layer for classification. Arrows indicate data flow; numbers below the blocks denote channel counts (and, where shown, spatial dimensions).

The input consists of the TT-SNR map: the $512 \times 256$ grayscale image described in Sec III. It first passes through a convolutional layer with eight filters and kernel size $(3 \times 3)$ and a stride of 2 (i.e., the step size of the convolution). This produces a $256 \times 128 \times 8$ output tensor, which is then normalized using batch normalization (BN) [59] and activated with a rectified linear unit (ReLU) [60]. Next, the data pass through three sequential residual blocks, which progressively extract higher-level features (see Fig. 3). After these blocks, a GAP [61] layer aggregates the most relevant information by averaging each feature map into a single value, producing a feature vector. This vector is then passed to a FC layer that outputs a classification score for the input image using a softmax function. To mitigate overfitting, a dropout layer with a dropout rate of $p = 0.3$ is applied between GAP and FC layers, randomly setting the chosen fraction of the input to zero during the training. The network is relatively lightweight, consisting of approximately 20,000 trainable parameters and is optimized using the Adam algorithm for 20 epochs. Adam adaptively adjusts learning rates based on estimates of the first and second moments of the gradients of the loss function [62].

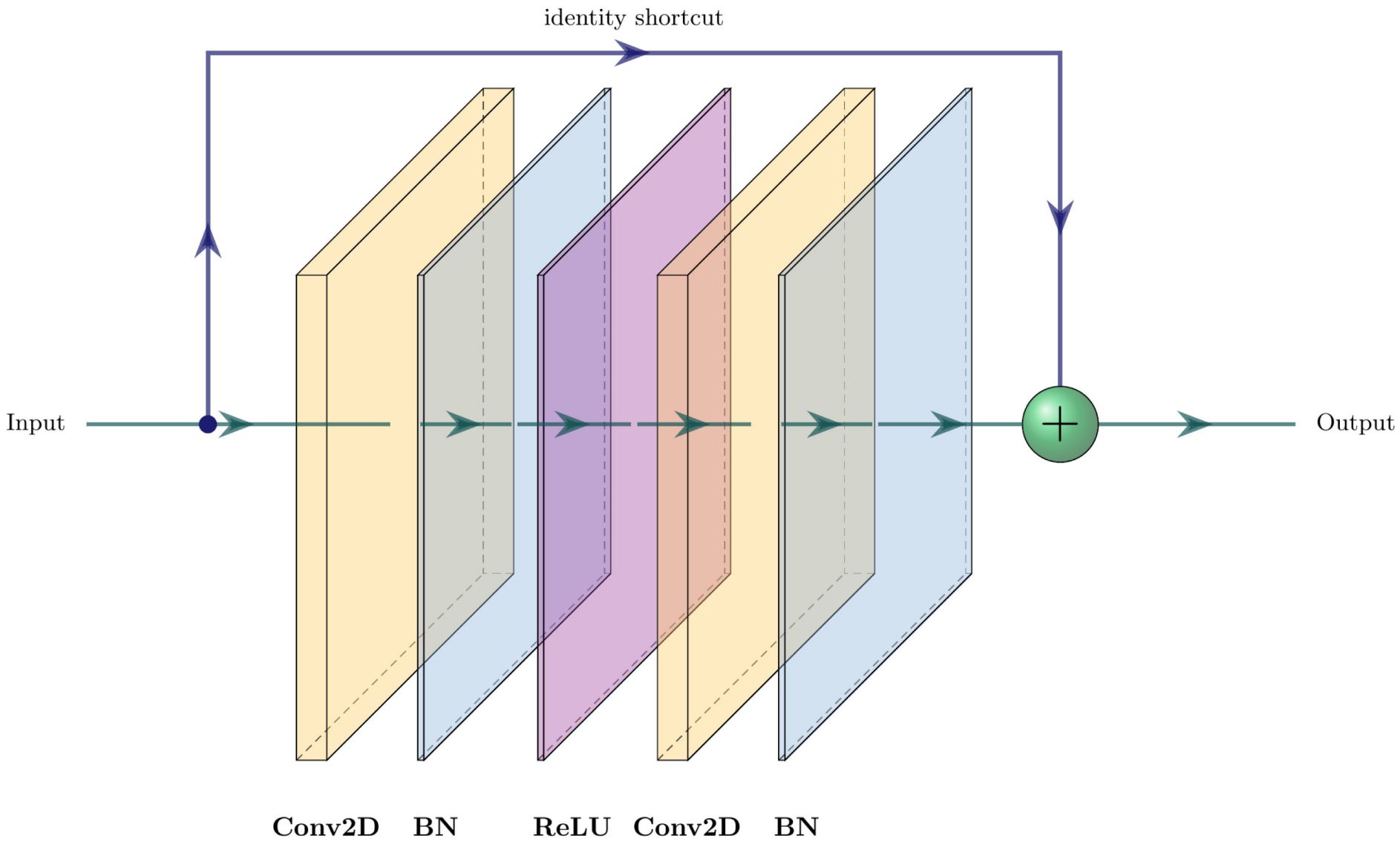

FIG. 3. Residual block. The lower arrow represents the input passing through a sequence of convolutions and normalizations. The upper one carries the identity. The green ⊕ denotes an elementwise addition. Arrow directions indicate data flow through the layers (Conv2D-BN-ReLU-Conv2D-BN in this example).

## V. SIMULATION 1

As a first dataset, we considered the idealized case of purely Gaussian noise with synthetic BNS signals injected in the data. By injection we mean the sum of the pure-noise data and a waveform generated from a set of parameters using an approximant. This dataset contains 74,800 images for each class, and the train-validation-test split follows the procedure described above. The relatively small size of the network allowed a training time of ∼8 hours on two Intel(R) Xeon(R) E5-2650 v4 @ 2.20 GHz CPUs, each providing 12 physical cores, for a total of 24 cores. The injected signals were generated by sampling the component masses uniformly ($\mathcal{U}$) in the interval $m_{1,2} \in \mathcal{U}[1.4 - 3]M_\odot$. No spins were assigned to the components. The distance distribution of the sources follows a power-law probability density distribution $\sim d^2$ over the range $d \in [50\text{–}1000]$ Mpc.

The metric used to assess the performance of the network is the cumulative distribution function (CDF) for the population $t$—noise (t = n) or injection (t = s)—defined as

$$\mathrm{CDF_t} = \frac{1}{\mathrm{N_t}} \sum_{i}^{\mathrm{N_t}} \theta(r_i - \hat{r}). \tag{9}$$

Here, $\mathrm{N_t}$ is the number of samples in the test dataset for the population $t$, $\theta$ denotes the Heaviside step function, and $r_i$ is the rank (or score) assigned by the network to the $i$-th sample. This score $r_i$ is compared to a fixed threshold $\hat{r}$; the CDF thus quantifies the fraction of events in population $t$ with scores above the threshold. In other words, it characterizes the score distribution of each population as a function of $r$. A well-trained network is expected to assign scores $r \sim 0$ to noise-labeled images and $r \sim 1$ to injection-labeled maps.

Figure 4 shows the results of the test dataset classification performed by the EasyResNet. In Fig. 4(a) the top panel shows, for the signal population, the correlation between distance, $\rho_{\mathrm{opt}}$, and the network score (color bar). On the bottom [Fig. 4(b)] the CDF shows that the network fails to completely separate the two populations; furthermore, for lower values ($r < 0.3$) the populations overlap significantly. This degeneracy is expected, as distant injections, or equivalently, signals with low optimal SNR $\rho_{\mathrm{opt}}$, cannot be reliably distinguished from the noise distribution: the TT-SNR maps are constructed exploiting the signal-to-noise-ratio time series, and it follows that the farther the signals are, the fainter they will be, and consequently the map will fade into Gaussian noise. The output layer uses a softmax activation, so that the ranking statistic $r$ is bounded to [0, 1] by construction. A bounded statistic can in principle place a floor on the accessible false-alarm probability: if a finite fraction of noise samples saturates at unity, no threshold can separate them. We explicitly verified that this does not happen in our simulations. Saturation to unity is observed solely for injection-labeled samples, while no noise-labeled sample saturates in any of the campaigns presented below, so that the two populations always remain fully ordered.

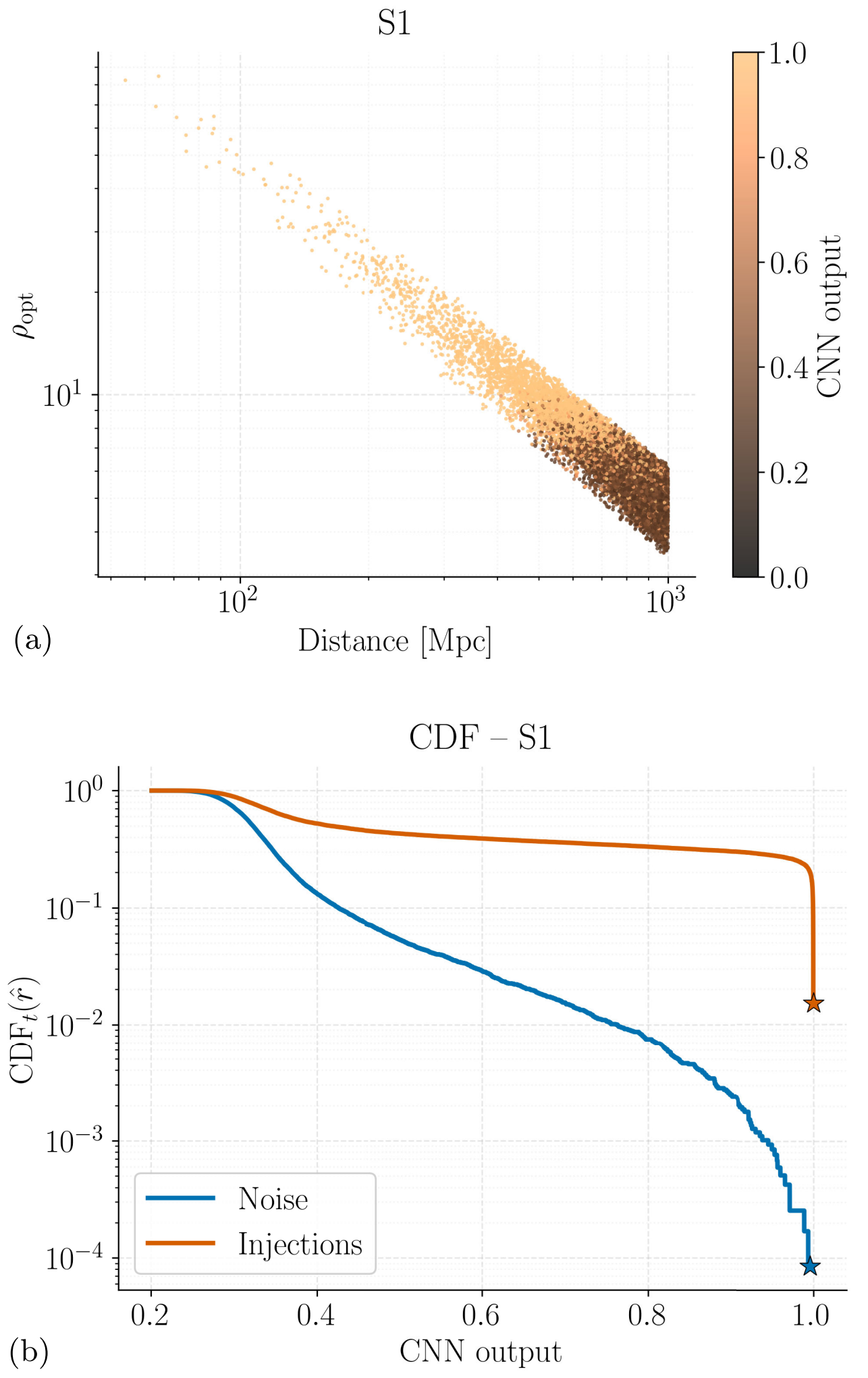

FIG. 4. Results for purely Gaussian noise and BNS injections. (a) Distribution of the injected signals in the $\rho_{\mathrm{opt}}$–distance plane, colored by CNN output. (b) CDFs for the noise and signal populations.

To compare the performance of the CNN with that of the $\rho(t)$ statistic, we collected the maximum value of the SNR time series ($\rho$) in the time interval considered to construct the TT-SNR maps for each data segment (4s). Then we used the receiver operating characteristic curve (ROC) as a metric. This curve is obtained by considering the false alarm probability (FAP) and the true alarm probability (TAP) as a function of a varying classification threshold. The FAP and TAP are defined as

$$\mathrm{FAP} = \mathrm{CDF}_{\mathrm{n}} = \frac{1}{N_{\mathrm{n}}} \sum_{i}^{N_{\mathrm{n}}} \theta(k_i - \hat{k}) \tag{10}$$

$$\mathrm{TAP} = \mathrm{CDF}_{\mathrm{s}} = \frac{1}{N_{\mathrm{s}}} \sum_{i}^{N_{\mathrm{s}}} \theta(k_i - \hat{k}). \tag{11}$$

Here $\mathrm{N_n}$ ($\mathrm{N_s}$) are the total number of noise (signal) samples in the test dataset. The variable $k$ denotes the statistic under consideration, either the matched-filter maximum $\rho^{\max}$, its reweighted counterpart $\rho^{\max}_{\mathrm{rw}}$ [Eq. (7)], or the CNN-output score $r$. The threshold $\hat{k}$ is varied to generate the ROC curve, which expresses the TAP as a function of the FAP. Results are reported in Fig. 5.

We see that the CNN statistic is less efficient in recovering signals with respect to the classical statistics $\rho$. For completeness, Fig. 5 also shows the reweighted statistic $\rho^{\max}_{\mathrm{rw}}$, which is discussed in detail in Sec. VI. In pure Gaussian noise there are no glitches for the $\chi^2$ test to reject, and $\rho^{\max}_{\mathrm{rw}}$ closely tracks $\rho^{\max}$, lying slightly below it: here the reweighting brings no benefit and can only down rank genuine signals that are imperfectly matched by the one-dimensional bank. This result is not surprising, since the residual network sees only an undersampled representation of the SNR time series: recall that we compressed the image into a $512 \times 256$ picture which impacts the representation of the TT-SNR map. The degradation relative to the matched-filtering statistic is a direct consequence of the compression described in Sec. III. To test whether it can be recovered by a different preprocessing, we regenerated the Simulation-1 dataset applying an RMS smearing of [33] to the matched-filter output prior to normalization. With this configuration the detection efficiency at false-alarm probability $10^{-4}$ decreases from 0.20 to 0.155, so the smearing does not close the gap with respect to $\rho$. The reason is that our compression factor is substantially larger than the one proposed in [33]: there, smearing is applied along the time axis alone, and all 256 templates are retained, whereas here the map is reduced along both the time and template axes. A like-for-like reproduction of that result would require a template bank small enough to avoid compression along the template direction altogether. Adopting such a bank would imply repeating the entire study from scratch, a cost in human rather than in computational terms, and would also entail a maximum mismatch considerably larger than the 3% assumed here, which is not a favorable regime in which to operate a matched-filtering search. This trade-off between the size of the bank and the amount of compression needed is nevertheless an interesting direction to investigate in future work. A further, related limitation is that the 4-second window captures only the portion of the TT-SNR map structure closest to the merger, whereas for BNS signals the response of the bank extends over a time comparable to the in-band duration of the templates, i.e., tens of seconds above 30 Hz. We nevertheless expect this truncation to play a minor role compared with the compression, since the central part of the butterfly-shaped structure is more visible and more discriminating than the tips of its wings, while enlarging the window would considerably increase the computational cost. We therefore regard the deficit in pure Gaussian noise as a limitation of the current implementation rather than of the TT-SNR map representation itself. The advantage of using such algorithms appears if we instead consider also glitches that affect the data stream, as is the case of the following sections.

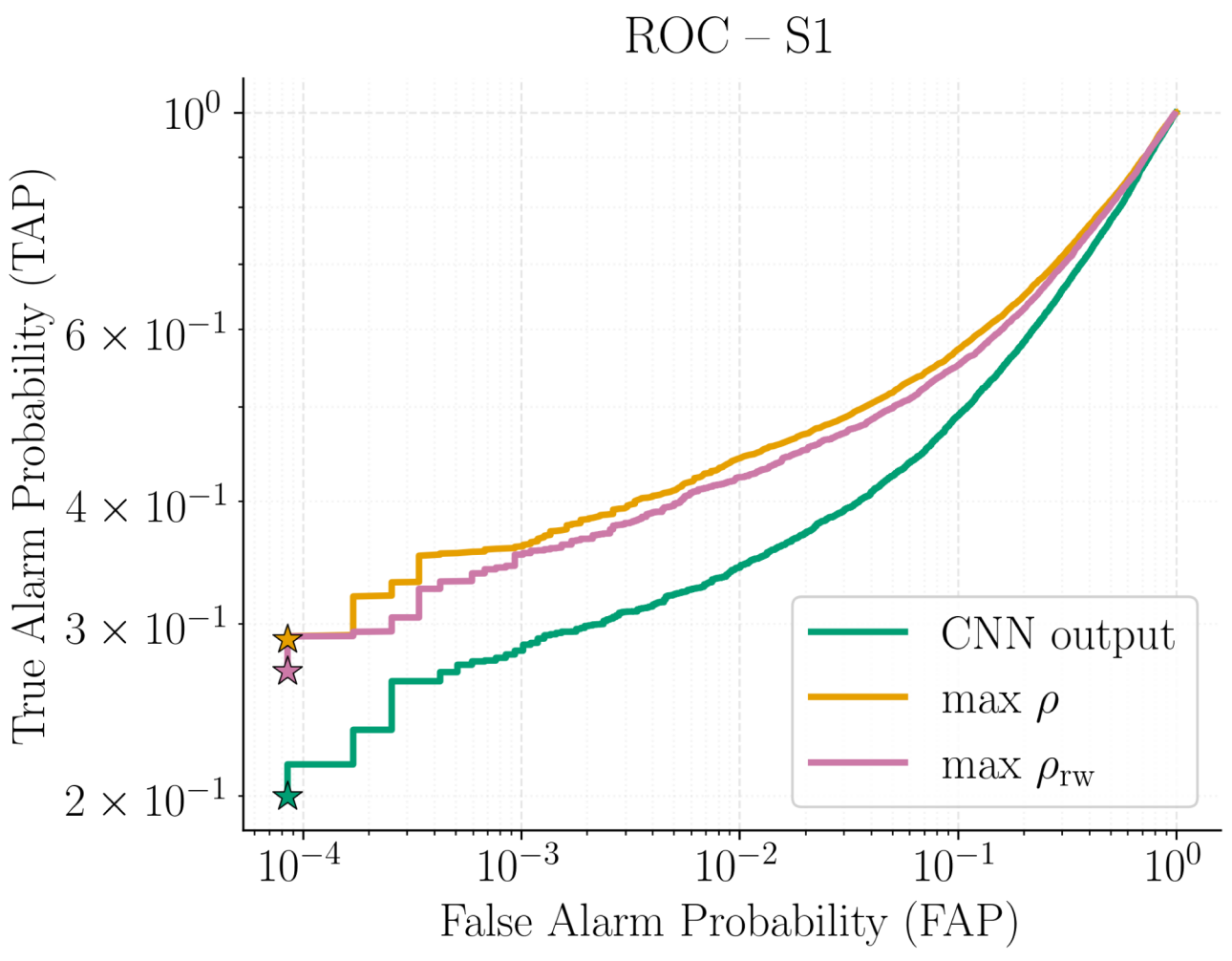

FIG. 5. Receiver operating characteristic curve considering injections in pure Gaussian noise, for the CNN output $r$, the matched-filter maximum $\rho^{\max}$, and the reweighted statistic $\rho^{\max}_{\mathrm{rw}}$.

## VI. SIMULATION 2

The second simulation considered in this work includes a dataset contaminated by non-Gaussian noise, which affects both the noise and injection classes. The dataset consists of 80,000 TT-SNR maps per class, divided into training-validation and test as previously described. In this case, the injected signals include both BNS and BBH systems, in equal proportions. The component masses for BNS and BBH injections were drawn from two different distributions. For the BNS systems, we retained a uniform distribution for the component masses as before. For BBH systems, the primary mass $m_1$ was sampled from an astrophysically motivated distribution [63], and modeled as a power law with slope $\alpha = 3.5$ and a Gaussian peak centered at $34M_\odot$. The secondary mass $m_2$ was computed via the mass ratio $q = m_2/m_1$, where $q \sim \mathcal{U}[0.15, 1]$, and the final sampling followed the transformation $q = q^{0.5}$ to favor symmetric systems. The allowed range for the primary mass was $m_1 \in [3\text{–}200]M_\odot$. The amplitudes of both BNS and BBH signals were scaled according to the chirp distance, defined as

$$d_{\text{chirp}} = \left(\frac{1.22}{M_c}\right)^{\frac{5}{6}} \cdot d, \qquad (12)$$

where $d$ is distributed according to a second-order power law, as in the previous case. This scaling accounts for the fact that the observable amplitude of a gravitational-wave signal depends on the chirp mass $M_c$, making $d_{\text{chirp}}$ a more appropriate parameter for distance scaling when comparing signals of different masses [64]. In this simulation, we injected sine-Gaussian glitches randomly into 30% of data segments, on both injection and noise labeled sets, to emulate non-Gaussian transient noise. Glitches were placed well outside the signal window as well, since the matched-filtering response to such transients is known to extend significantly beyond the glitch time itself (see [65]). A sine-Gaussian in time domain can be written as

$$g(t) = Ae^{\left(\frac{-(t-t_0)^2}{\tau^2}\right)} \cos(2\pi f_0 t + \phi_0). \qquad (13)$$

In this expression, $A$ denotes the amplitude, and $\tau = \frac{Q}{2\pi f_0}$ the duration of the pulse, where $Q$ is the dimensionless quality factor. The parameters $f_0$ and $t_0$ specify the central frequency and central time of the sine-Gaussian, respectively, while $\phi_0$ is the phase at $t = 0$. We injected the sine-Gaussian glitches uniformly in a time window $\Delta t \in [-63, 0.5]$ s, taking the true merger time of the binary system at $t = 0$. This simulates the possible effect of the glitches even when they occur well before the actual signal. The remaining sine-Gaussian parameters were sampled from uniform distributions $\mathcal{U}$, with the corresponding ranges listed in Table I.

In this simulation, we compared the CNN-assigned score $r$ with an effective statistic that incorporates the $\chi^2$ test as is commonly done in the PyCBC pipeline to suppress glitch-induced SNR. The procedure involves computing the $\chi^2$ discriminator over a given number of frequency sub-bands $N_b$, according to Eq. (6). As already pointed out in Sec. II, $N_b$ should in general be chosen as a function of the template; here, for the practical reasons discussed there, we fix it to $N_b = 16$ for all the templates in the bank. The reweighted statistic obtained is the $\rho_{\text{rw}}$ as described in Eq. (7).

We then identify the maximum value $\rho_{\text{rw}}^{\text{max}}$ within the 4-second time window and compare it to the score $r$ for the corresponding TT-SNR map. To evaluate and compare the classification performance of the CNN and the traditional effective statistics, we again use the ROC curve.

We expect that the reweighted statistic $\rho_{\text{rw}}^{\text{max}}$, which incorporates the $\chi_r^2$ test, will be significantly more effective in separating noise from signal than the raw matched-filter maximum $\rho^{\text{max}}$. We then compare the ROC curves obtained using $\rho_{\text{rw}}^{\text{max}}$ and the CNN output $r$ statistics to quantify the network's ability to distinguish between glitches and true

TABLE I. Parameter ranges used for the injected sine-Gaussian signals.

| Parameter | Value range |
|---|---|
| Amplitude $A$ | $[1, 9] \times 10^{-22}$ |
| Central frequency $f_0$ | $[20, 2048]$ Hz |
| Quality factor $Q$ | [3, 400] |
| Initial phase $\phi_0$ | [0, 6] |

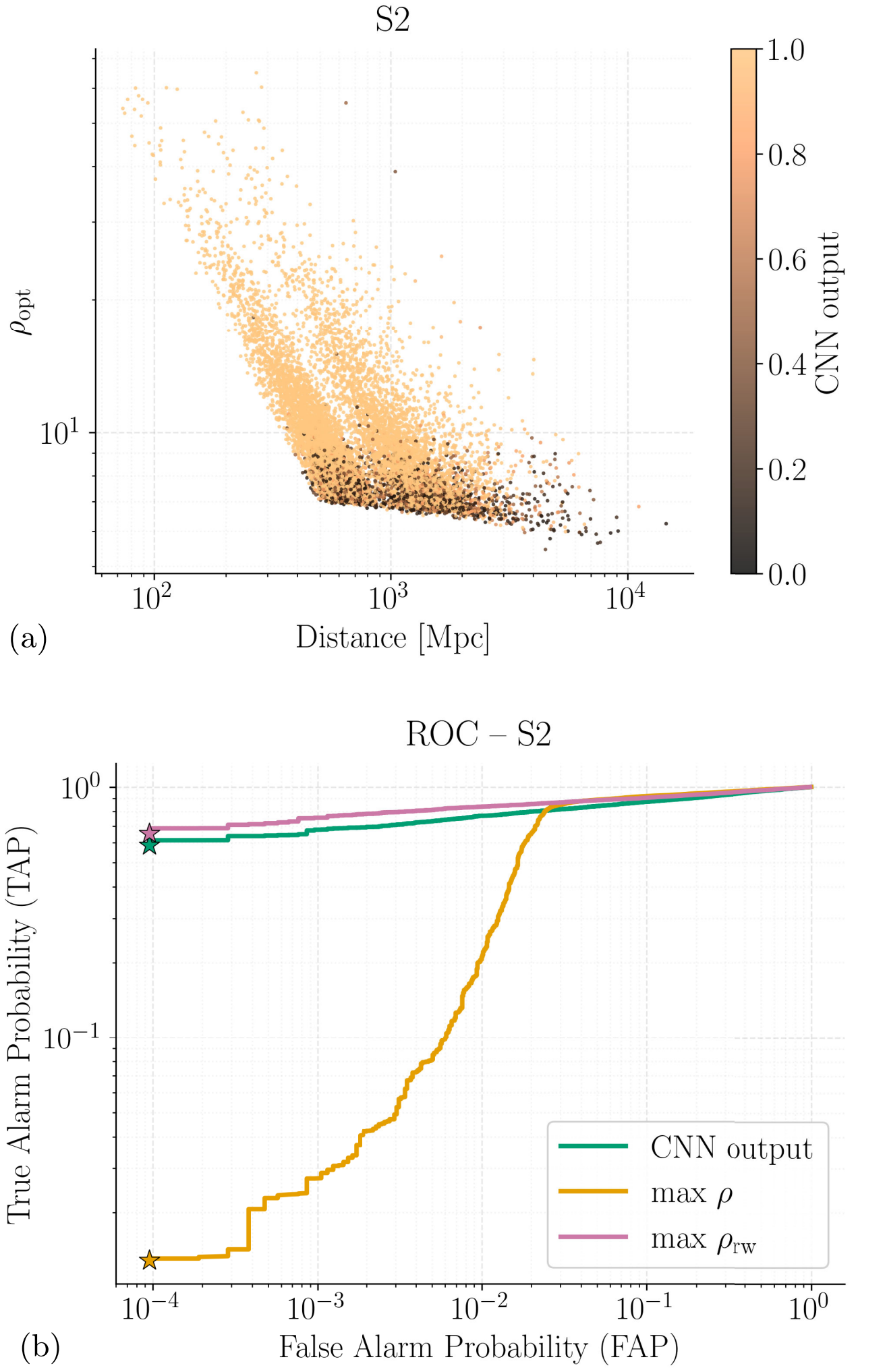

FIG. 6. Results for BBH and BNS signals injected in Gaussian noise polluted with sine-Gaussian glitches. (a) Distribution of the injected signals in the $\rho_{\text{opt}}$–chirp distance ($d_{\text{chirp}}$) plane, colored by CNN output. (b) ROC curves for the $\rho_{\text{rw}}^{\text{max}}$, CNN-output $r$, and $\rho^{\text{max}}$ statistics.

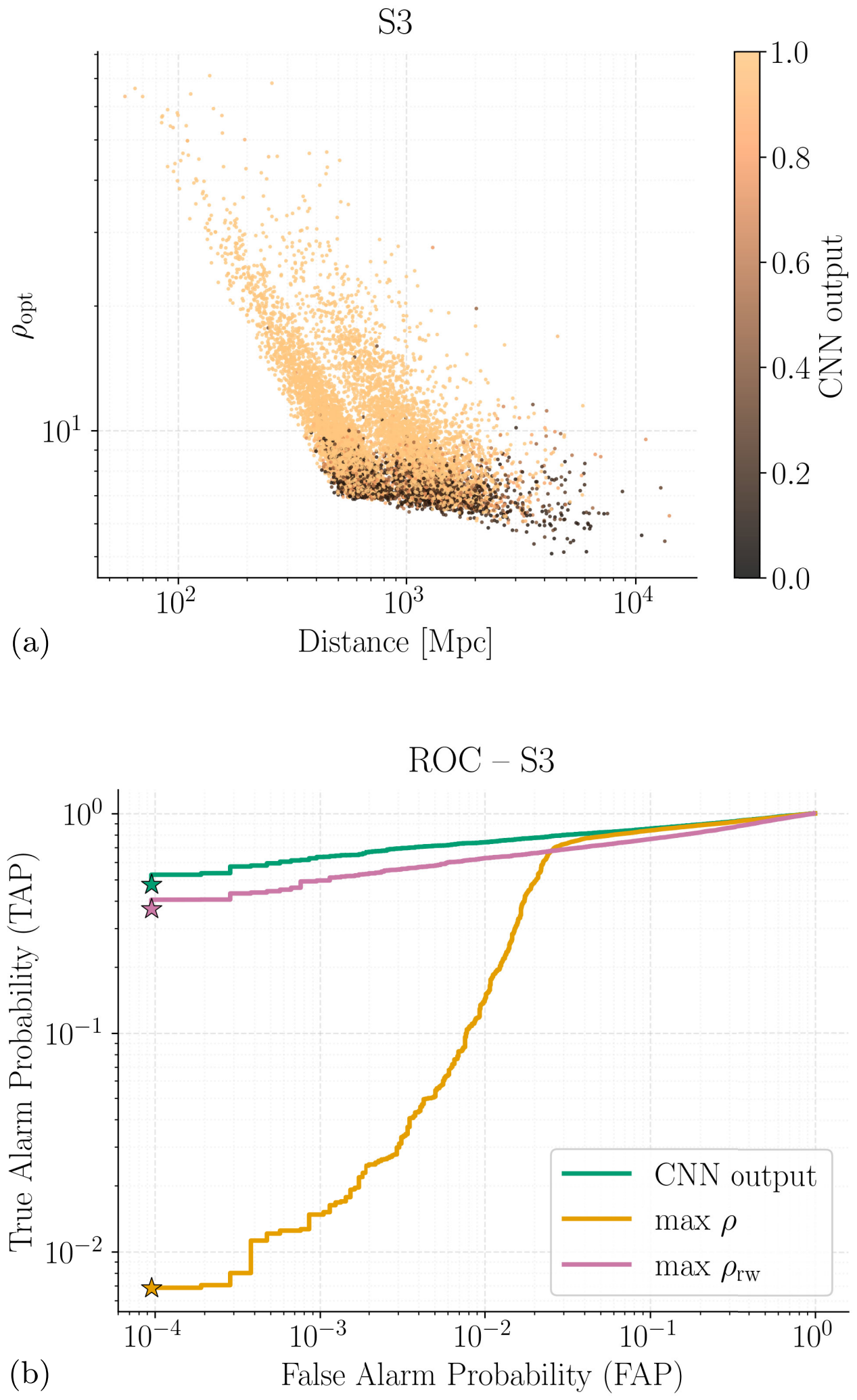

FIG. 7. Results for BBH and BNS signals encoding spin parameters and injected in Gaussian noise polluted with sine-Gaussian glitches. (a) Distribution of the injected signals in the $\rho_{opt}$–chirp distance ($d_{chirp}$) plane, colored by CNN output. (b) ROC curves for the $\rho_{rw}^{max}$, CNN.output $r$, and $\rho^{max}$ statistics.

signals relative to the classical approach. Results are reported in Fig. 6(b).

In the results we observe that the CNN output outperforms the raw matched-filter statistic $\rho$ for low false alarm probabilities and obtains results compatible with the reweighted statistic $\rho_{rw}^{max}$, suggesting that the network is capable of separating glitches and signals as efficiently as the statistics that exploits the $\chi^2$ test veto. Even in this case, we see that the louder the signals are, the better the CNN is able to detect these [Fig. 6(a)].

## VII. SIMULATION 3

In this study, we included additional physical information in the injected signals that is not represented in the template bank. Specifically, spin effects were incorporated by assigning aligned-spin components drawn from uniform distributions. For neutron stars, we used a conservative spin range of $\chi_{z,NS}^{1,2} \in \mathcal{U}[-0.5, 0.5]$, while for black holes we adopted $\chi_{z,BH}^{1,2} \in \mathcal{U}[-0.9, 0.9]$, reflecting the typically higher spin magnitudes of BHs [66]. The other parameters, such as chirp distance and mass distributions, are the same as those used in the simulation-2 campaign. As in the previous studies, the dataset was divided into training-validation-test in a 70 − 15 − 15 split. The total dataset consists of 80,000 TT-SNR maps per class for both noise and injection samples. The performance comparison followed the same procedure as in the previous section, and is represented in Fig. 7(b).

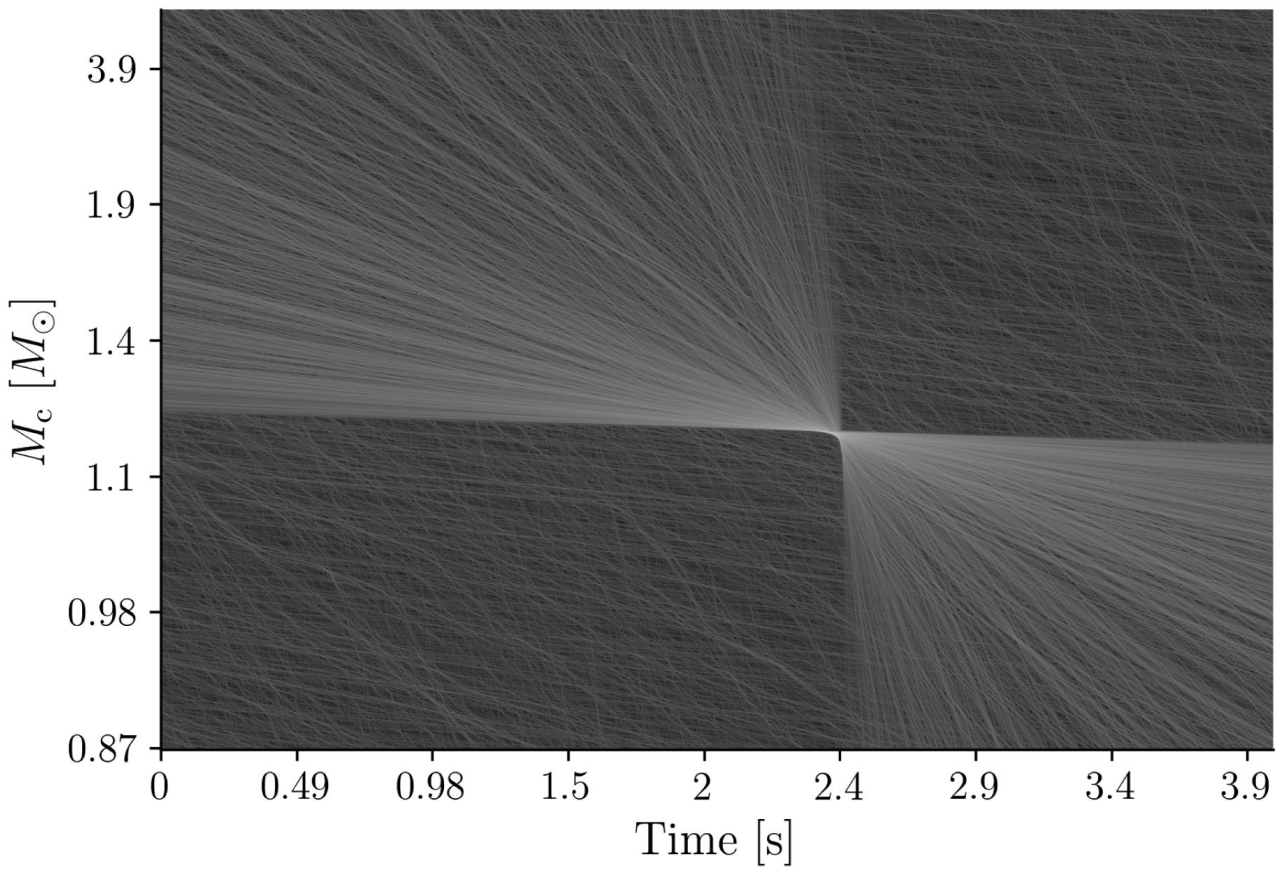

FIG. 8. BNS spinning ($\chi_1^z = \chi_2^z = 0.5$; 1.4–1.4$M_\odot$) signal. The TT-SNR map shows the presence of spin by this "rounding" angle effect in the lower-left of the butterfly structure.

In this case the physical effect of spin absence from the template bank causes the $\chi^2$ test to down rank the SNR associated with the true injections. In contrast, EasyResNet is not affected by this mismatch and achieves performance comparable to that of the matched-filtering search, even in the presence of spin. The imprint left by the spins on the map itself is illustrated in Fig. 8, which shows the TT-SNR map of a 1.4–1.4$M_\odot$ BNS with aligned spins $\chi_1^z = \chi_2^z = 0.5$: the presence of spin manifests as a characteristic "rounding" of the angles in the lower-left region of the butterfly-shaped structure.

## VIII. SIMULATION 4

In this final simulation, we extend the investigation of the capability of the TT-SNR map to encode physical information that goes beyond the physics explicitly included in the template bank used to construct the map itself. In particular, we expect this representation to capture not only the presence of a signal, but also the specific physical effect characterizing the injected waveform. Moreover, motivated by the encouraging results obtained in the previous section, we anticipate that a network trained on these TT-SNR maps will achieve performance comparable to—or potentially exceeding—that of the classical $\rho_{\rm rw}$ method. To this end, we considered five different families of signal injections, each characterized by a different physical effect or scenario that the TT-SNR map is expected to represent and that the network is trained to distinguish. In this study we deliberately focus on isolating the impact of individual physical effects rather than reproducing their expected astrophysical distributions. Consequently, we adopt nonphysical distributions for these effects, which nevertheless suffice for the exploratory nature of this analysis. However, the injected distances still follow the chirp-distance distribution, to assess the network's ability in recovering signals with diverse physics in a loud-to-faint environment. First, we consider an extreme version of the scenario described in the simulation 3 campaign, restricting the spin configurations to purely aligned or antialigned cases with respect to the orbital angular momentum. Specifically, we adopt extremal spin values for the BNS ($\chi^z_{1,2} = \pm 0.5$) and for the BBH ($\chi^z_{1,2} = \pm 1$) population. The mass distributions of injected signals are identical to those used in the simulation 3 campaign.

The second family of simulations consists of superimposed signal injections. This scenario is of particular interest in the context of the future 3G detectors such as the Einstein Telescope [16] and Cosmic Explorer [17]. Owing to their increased sensitivity and extended low-frequency coverage, these instruments are expected to observe a high rate of loud signals, which will likely overlap in the detector strain data. Analyzing such data streams poses significant challenges, and the method explored here may provide a valuable approach in this direction [67]. In this study, we consider only BNS signals with component masses and spins drawn from uniform distributions: $m_{1,2} \sim \mathcal{U}(1.4, 3) M_\odot$; $\chi^z_{1,2} \sim \mathcal{U}(-0.5, 0.5)$. For simplicity, we inject only two signals per realization, separated by a short temporal delay uniformly distributed in the interval $[-0.02, 0.02]$ s. The second injection is constructed to have component masses close to the first signal, in order to mimic the most challenging case of highly similar overlapping events. Specifically, the masses of the second signal are defined as $m'_{1,2} = m_{1,2} + \mathcal{U}(-0.02, 0.02) M_\odot$. A representative TT-SNR map from this family, together with an enlarged view around the merger time, is shown in Fig. 9. A clarification is needed on what constitutes a detection for this family. The analysis performed here is a segment-level binary classification: each 4-second window is assigned a single score, and the segment is labeled as detected if the corresponding statistic exceeds the threshold, irrespective of how many signals it contains. The same definition is applied to the reference statistics, for which we take the maximum of $\rho$ and $\rho_{\rm rw}$ over the window, so that the comparison between the three rankings is performed on identical terms. No attempt is made to resolve, count, or separately characterize the individual components of an overlapping pair. Resolving the multiplicity of overlapping events is a distinct problem, addressed for instance in [67], and lies outside the scope of this proof of concept. What the present test measures is whether the morphology imprinted on the TT-SNR map by two closely spaced, nearly identical signals remains recognizable as signal-like.

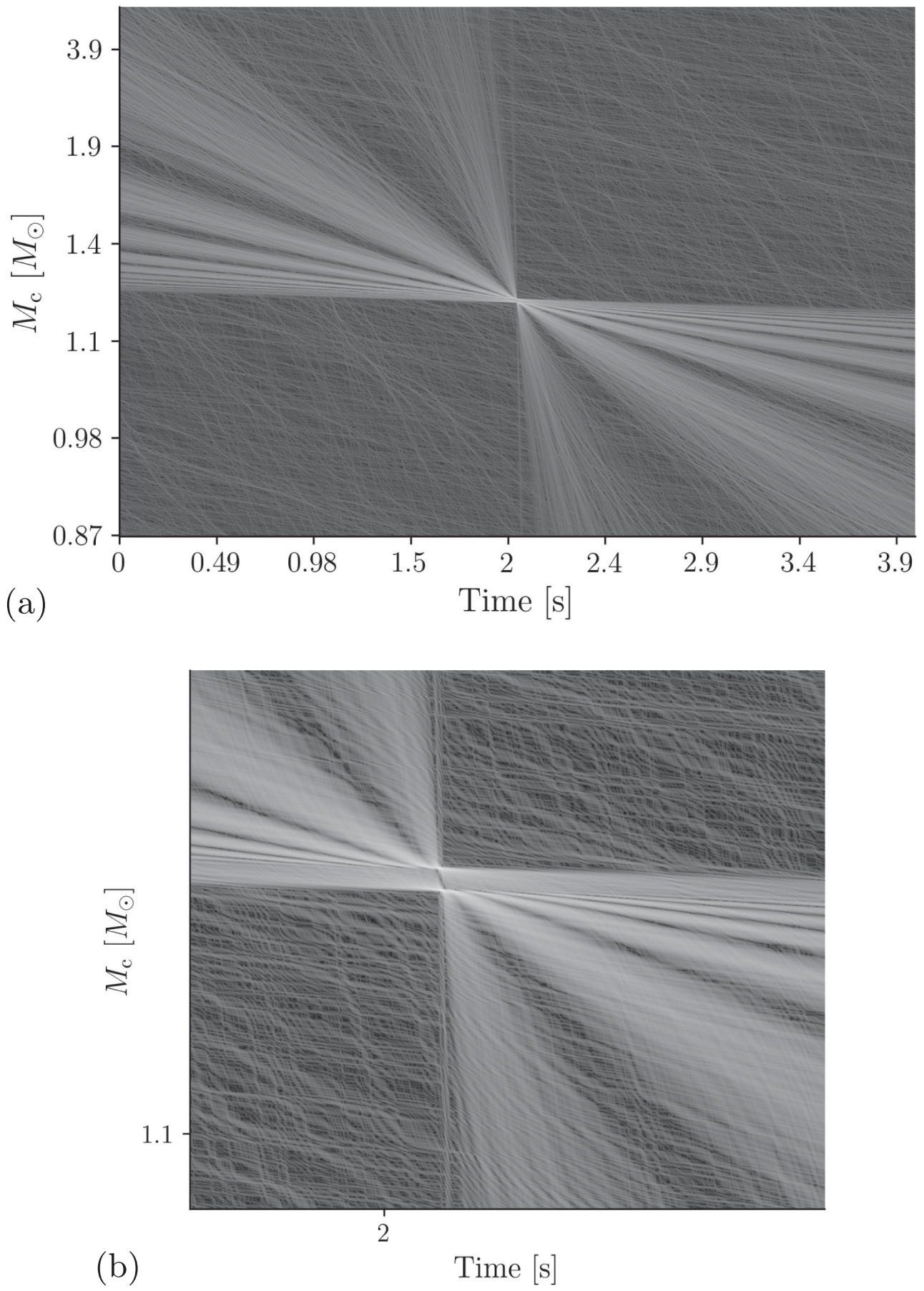

FIG. 9. TT-SNR map resulting from injected nonspinning BNS (1.4–1.4$M_\odot$) superimposed signals. (a) Full map: it clearly encodes the presence of two signals closely separated in time and nearly identical in parameter space, and an interference-like pattern emerges from the superposition of the two signal representations. (b) Zoomed view around the merger time.

The following injection campaign incorporates the presence of higher-order modes (HoMs) [68] in the injected signals (see Fig. 10). This case is particularly relevant because higher-order modes encode a more accurate description of the gravitational waveform, arising from additional contributions in the spherical-harmonic decomposition of the solution to the Einstein equations. To maximize the impact of HoMs on the observed waveform, all the signals are injected with an inclination angle $\iota = \frac{\pi}{2}$, corresponding to an edge-on orientation of the binary system. In this campaign, we consider nonspinning BNS and BBH systems, with component masses drawn from the same uniform and power-law distributions adopted for the two populations in the previous simulations. The waveforms are generated using the SEOBNRv4HM approximant [69,70].

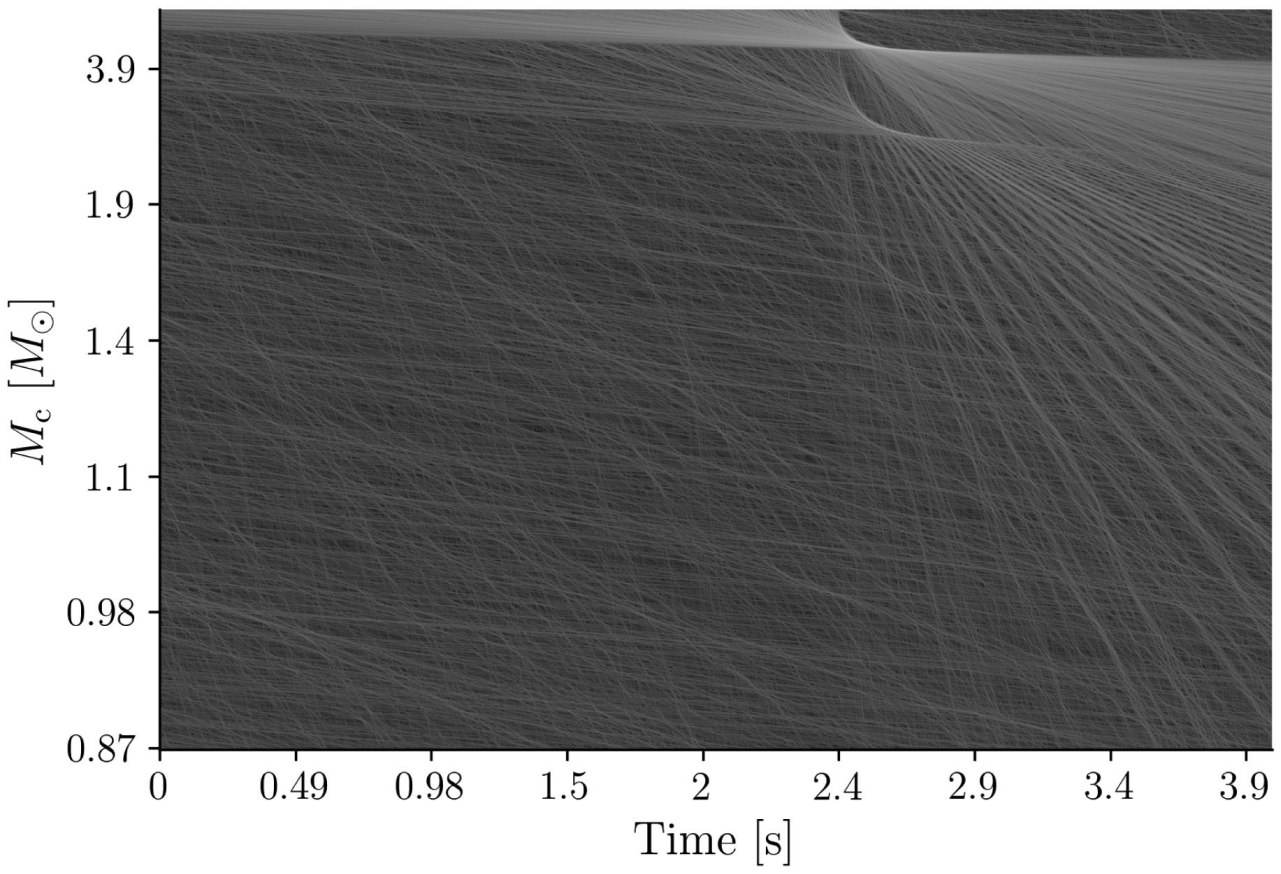

FIG. 10. Example of a NSBH ($1.4$–$25M_\odot$) waveform including higher-order modes, injected with inclination $\iota = \frac{\pi}{2}$. The TT-SNR map shows a clear modification of the central region, where the sharp rectangular features typically present in the standard representation are smoothed out by the contribution of higher-order modes. A similar structure is also visible at lower amplitudes, repeating regularly at later times.

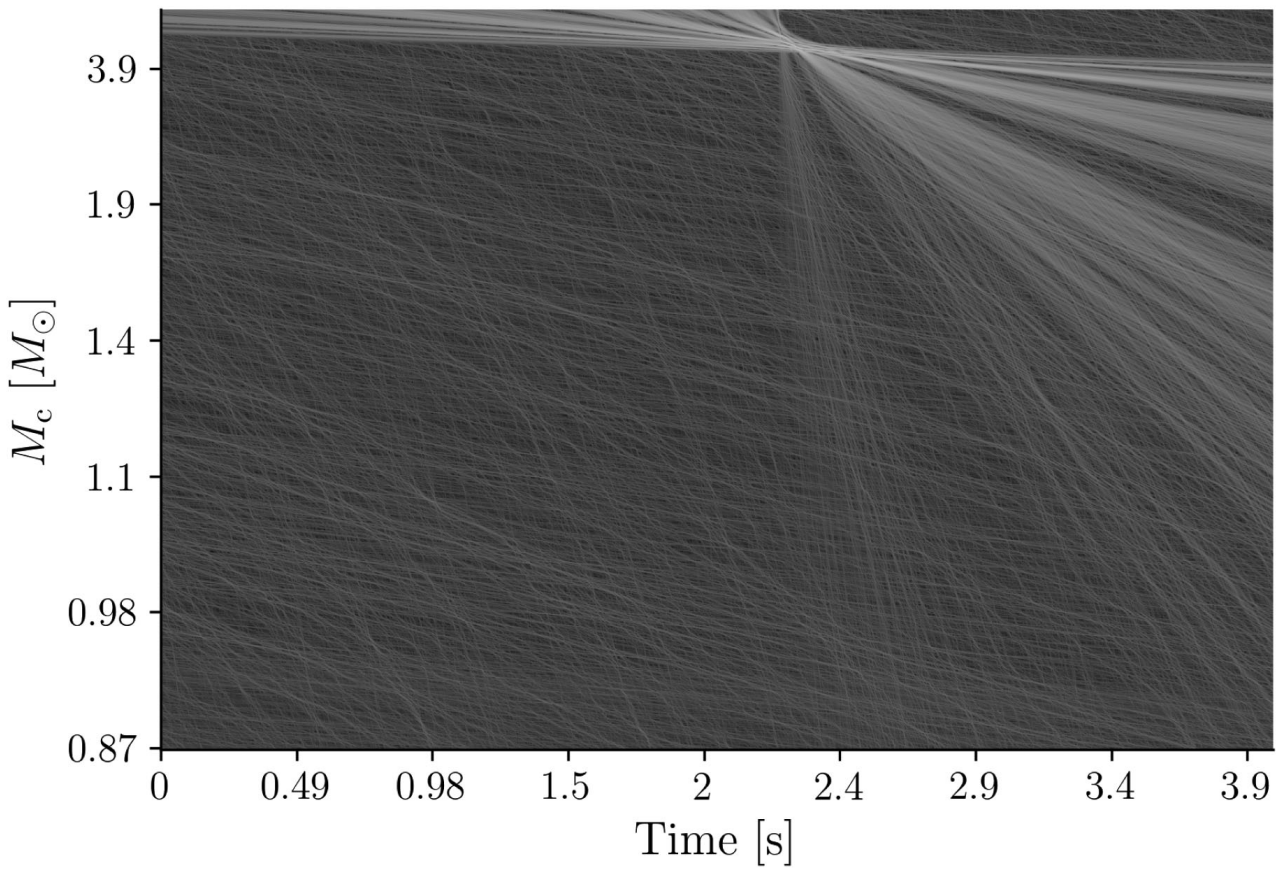

FIG. 11. TT-SNR map of precessing NSBH system ($1.4$–$25M_\odot$) injected with inclination $\iota = \frac{\pi}{2}$. The effect of spin-induced precession manifests as a modulation of the TT-SNR map structure, producing beamlike features radiating from the center of the map. These patterns reflect the time-dependent orbital-plane precession imprinted on the signal morphology.

The fourth family of injections investigates the impact of spin-induced precession on the TT-SNR map. Precessing signals are of particular interest because they encode rich information about the properties of the compact binaries, including the orbital geometry and characteristic length scales of the system [71,72]. Moreover, precession provides key insight into the spin distribution of the BBH population, thereby constraining their possible formation channels [73,74]. In this campaign, we considered a population of signals equally divided among BNS, NSBH, and BBH systems. The mass distributions for the BNS and BBH populations follow those adopted in the previous simulations. For the NSBH systems, the component masses are distributed as $m_1 \sim \mathcal{U}(10, 25)M_\odot$ and $m_2 \sim \mathcal{U}(1.4, 3)M_\odot$. For all three populations, the spin components are configured to induce strong precession by drawing the in plane components from uniform distributions, $\chi^x_{1,2}, \chi^y_{1,2} \sim \mathrm{U}(0.4, 0.5)$, while setting the aligned component to zero, $\chi^z_{1,2} = 0$. As in the previous injection families, the inclination angle is fixed to $\iota = \frac{\pi}{2}$ to enhance the precessional effects. Only the dominant quadrupole mode $(\ell, m) = (2, 2)$ is retained in this analysis. The approximant used is SEOBNRv5PHM [75]. An example of a precessing signal in this configuration is shown in Fig. 11.

The final injection campaign explores the impact of orbital eccentricity on the TT-SNR map representation. Eccentric gravitational-wave signals are of particular interest because they provide valuable clues about the formation channels of compact binaries, especially in dynamical environments such as globular clusters or galactic nuclei, where binaries may retain significant eccentricity when entering the detector sensitivity band [76,77].

This study considers nonspinning BNS and BBH signals. For the two populations, the orbital eccentricity is drawn from uniform distributions with relatively high values, chosen to enhance the imprint of eccentricity on the waveform morphology: $e \sim \mathcal{U}(0.2, 0.5)$. Such values are intentionally higher than those expected for the majority of astrophysical sources detectable by current instruments, but they are well suited for this exploratory study aimed at testing the network's sensitivity to eccentric features that are not encoded in the template bank. Waveforms are generated using the SEOBNRv5E_td approximant [78]. A representative example of an eccentric injection from this campaign is shown in Fig. 12.

For all injection families, 30% of the signals are contaminated with sine-Gaussian glitches. Each family also consists of 16,000 injections, resulting in a total of 80,000 TT-SNR map images labeled as injection. Following the same methodology adopted in the previous simulation campaigns, we construct a balanced dataset by including an additional 80,000 noise-labeled TT-SNR maps, generated from Gaussian noise and contaminated with the same ratio of sine-Gaussian glitches. The full dataset is then split into training, validation, and test subsets using the same proportions adopted for the previous populations $(70\% - 15\% - 15\%)$. The trained network is subsequently evaluated and compared against classical detection statistics through the ROC curves. Owing to the network's rapid convergence and strong learning capability, the number of training epochs is reduced to five, beyond which no statistically significant performance improvements are observed. The corresponding results are shown in Fig. 13. The results for each injection family are reported in Fig. 14.

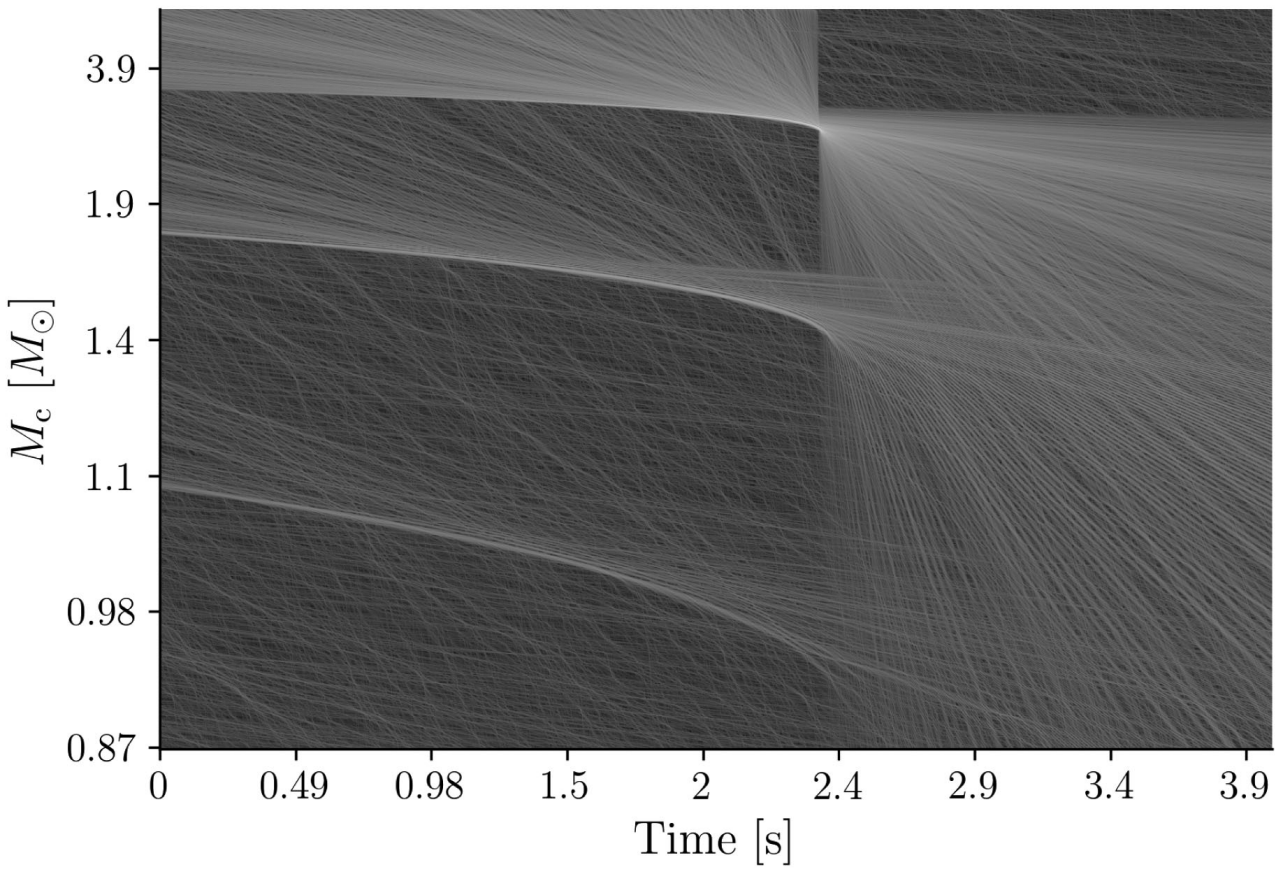

FIG. 12. Representative example illustrating the effect of eccentricity in the TT-SNR map. Here is shown a 1.4–6$M_\odot$ nonspinning system with $e = 0.3$. The imprint of eccentricity is clearly visible as a repeating structure beneath the main TT-SNR map pattern, appearing twice and introducing distinctive features associated with the noncircular orbital dynamics.

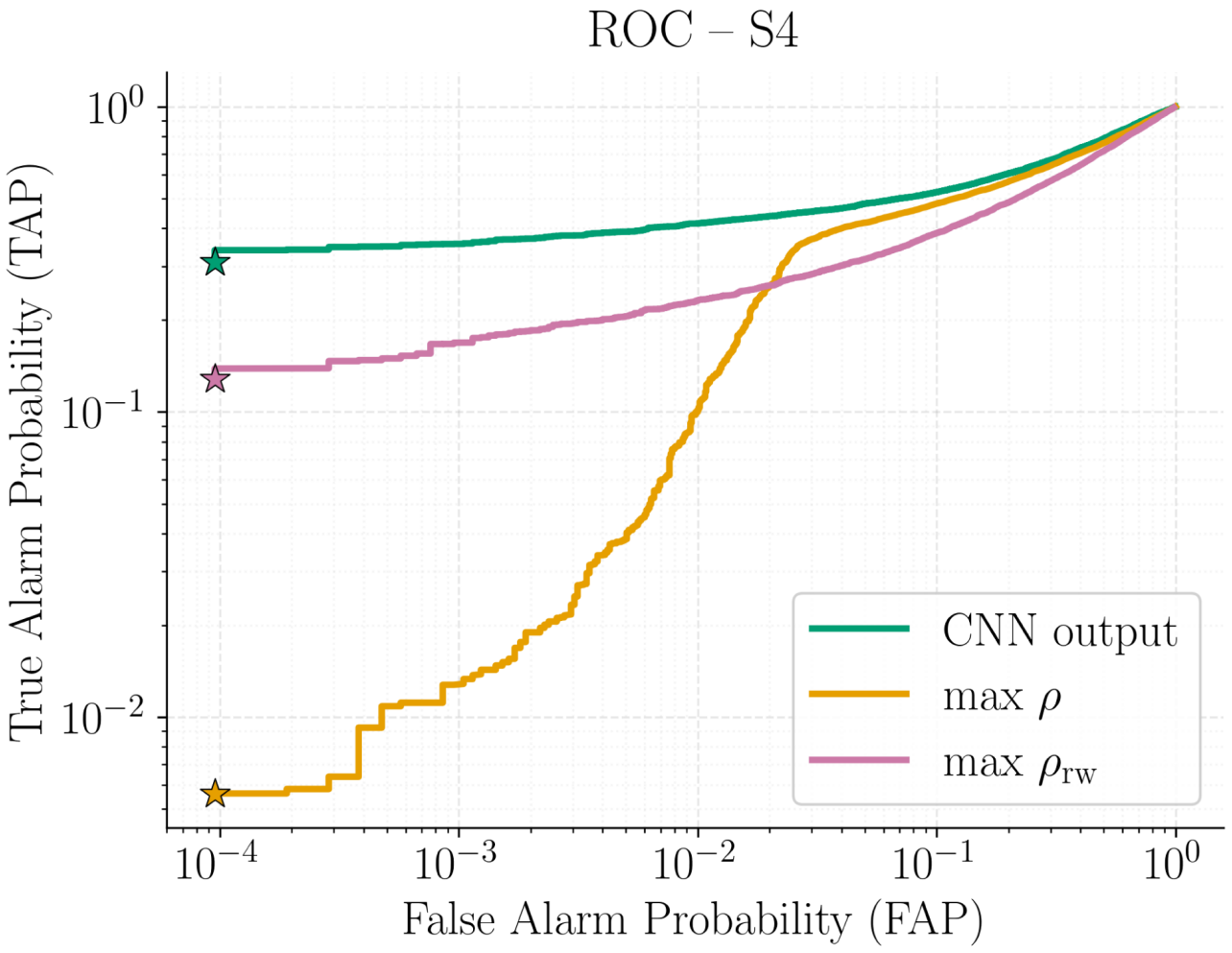

FIG. 13. Results of the ROC curves for the Simulation-4 injection campaign encoding physics not included in the template bank used to construct the TT-SNR map. The signals are injected in Gaussian noise polluted with sine-Gaussian glitches.

When the injection campaign is considered as a whole, the network exhibits an improved capability to discriminate signal from noise with respect to the previous simulations. The ROC curve obtained with the ResNet architecture lies above those of both the $\rho$ and $\rho_{\rm rw}$ statistics, with the margin widening toward low false-alarm probabilities: at $FAP = 10^{-4}$ the network recovers TAP ~ 0.30 against ~0.13 for $\rho_{\rm rw}$ and $\sim 6 \times 10^{-3}$ for $\rho$. A systematic analysis of the individual injection families, however, reveals a markedly diverse behavior; see Fig. 14. The network efficiency at fixed FAP closely follows the loudness of each set: superimposed, whose injections populate the high-$\rho_{\rm opt}$ region of the distance-SNR plane, is recovered almost completely, followed by the extreme spins and eccentricity, while precessing and HoM sit at the bottom of both rankings. This ordering is a direct consequence of the detection mechanism: the network is trained on the morphology of the aggregated SNR in the TT map, so its sensitivity inherits the amplitude of the underlying matched-filtering response. The eccentricity case is the most favorable for the network, and it is also where the limitations of the reweighted statistic are most evident. Since the template bank contains no eccentric waveforms, the residual mismatch is redistributed into the $\chi^2$ statistic, and $\rho_{\rm rw}$ is penalized. At FAP $> 2 \times 10^{-2}$ the reweighted curve falls below the plain $\rho$ one: the veto aggressively suppresses genuine signals whose physics is not represented in the templates. The network, which is not subject to any such consistency test and exploits the time-template structure of the SNR map rather than its peak value alone, is not affected by this penalty. The precessing and HoM families are instead the hardest for all three statistics. With the inclination held fixed at $\iota = \frac{\pi}{2}$, the cross polarization vanishes, and the interferometers observe a single, linearly polarized combination; the dominant $\ell = |m| = 2$ amplitude is at its minimum, while subdominant modes and precession-induced amplitude and phase modulations are maximally excited. These injections are therefore simultaneously intrinsically quiet and maximally mismatched with respect to the nonspinning, quadrupolar bank, so that both $\rho$ and the TT-SNR map built from it are degraded; the coarse-grained representation produced by the compression procedure further reduces the amount of recoverable structure. In the precessing case, the network still provides the best performance of the three statistics (TAP ~ 0.095 against 0.07 and 0.01 for $\rho_{\rm rw}$ and $\rho$, respectively, at the lowest FAP probed for this family, FAP $\approx 2 \times 10^{-3}$, marked by the stars in Fig. 14), although at low absolute efficiency, whereas HoMs is the only configuration in which $\rho_{\rm rw}$ marginally exceeds the network over most of the FAP range. We stress that the per-family comparison is not fully homogeneous: the injection sets span different distance and optimal-SNR distributions, so part of the observed spread reflects the properties of the campaigns themselves rather than the sensitivity of the algorithms. Taken together, these results indicate that the network recovers signals efficiently whenever they are sufficiently loud, and that its advantage is largest precisely where the injected physics is absent from the template bank and the $\chi^2$-based reweighting becomes counterproductive.

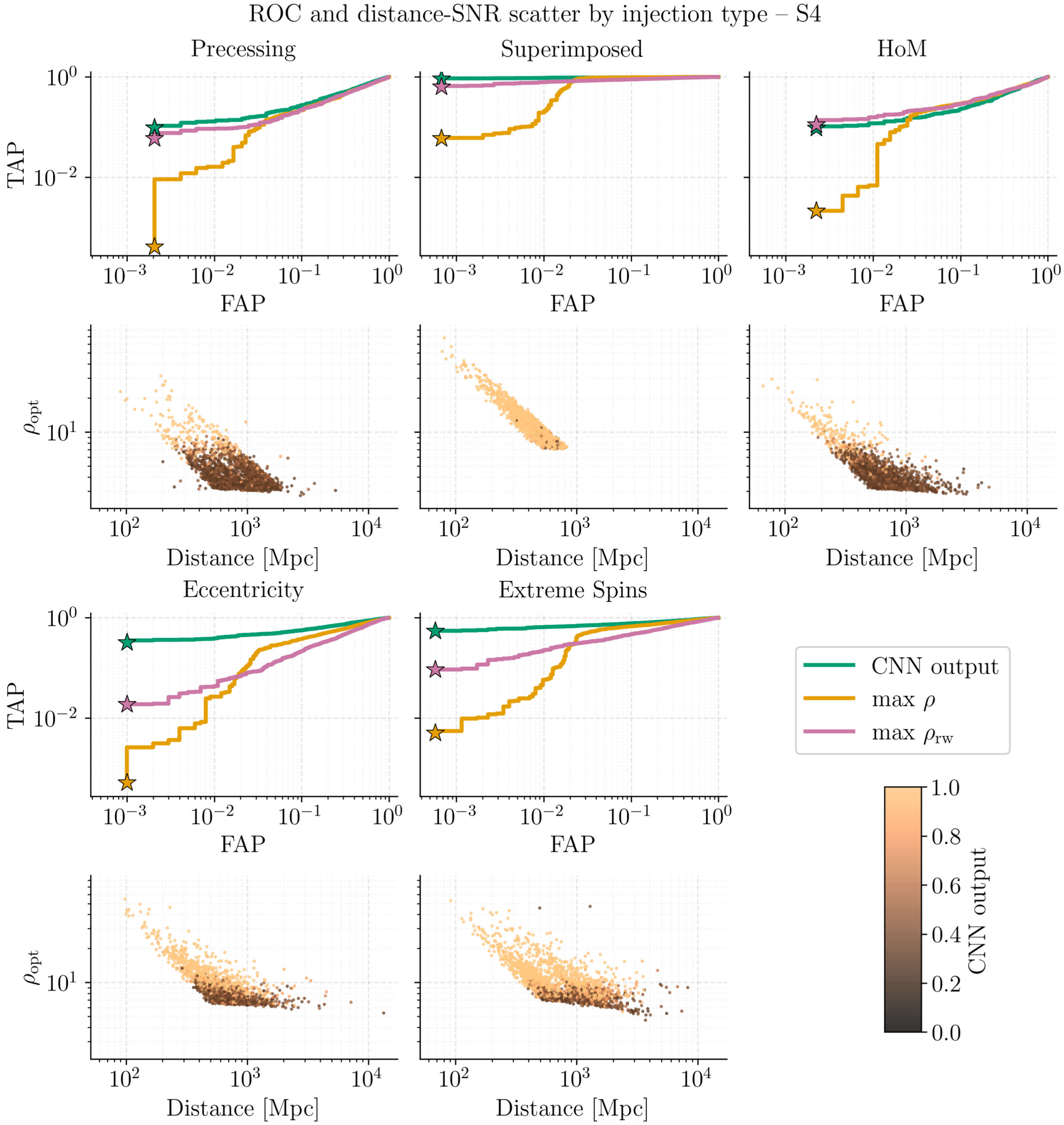

FIG. 14. Results for the five families of simulated signals: precessing, superimposed and HoM (first and second row), eccentricity and extreme spins (third and fourth row). For each family, the upper panel shows the ROC curves, and the lower panel the $\rho_{\rm opt}$–distance scatter plot, colored by the CNN output score. All ROC panels share the same axis limits, as do all scatter panels, so that the families are directly comparable.

## IX. CONCLUSIONS

In this proof-of-concept study, we trained and tested a convolutional neural network (EasyResNet) on TT-SNR maps constructed from matched-filtering SNR time series. Four successive simulation campaigns were performed. The first used injections drawn from a well-defined parameter space embedded in purely Gaussian noise. The second introduced sine-Gaussian glitches into the background and extended the injections to include BBH systems. The third and fourth incorporated physical effects—aligned component spins, higher-order-modes, precession, eccentricity, and overlapping signals—not represented at all in the template bank, in order to test the network's robustness relative to a matched-filtering analysis.

EasyResNet consistently learned the TT-SNR map patterns associated with the true signals and distinguished them from noise. The network's advantage over the classical ranking was most pronounced when the injected signals contained physical parameters not covered by the bank, a regime in which the classical matched-filter ranking is degraded. An advantage of this method is that the $\chi^2$ veto need not be computed for rejecting the glitches, since the network can learn automatically how signals and spurious noise behave in the TT-SNR map. This comes with fast inference—∼3 ms per image using only off-the-shelf CPUs without particular optimizations—making the method attractive for low-latency applications.

Since in this work the search is performed solely in terms of the chirp mass $M_c$, the resulting template bank is considerably reduced in size. Further studies are therefore required to confirm the detection sensitivity of the network relative to state-of-the-art CBC search pipelines that exploit full-sized template bank. If confirmed, this approach could offer a viable alternative for third-generation detectors such as the Einstein Telescope: a TT-SNR map-based analysis combined with a trained network could produce triggers using a significantly reduced template bank.

In this work, these results were obtained using standard CPU resources, without exploiting GPU-based parallelization for network training. This limitation necessitated the adoption of undersampled TT-SNR map representations, which negatively impacted the network's capabilities and overall performance. Access to greater computational resources in future studies is expected to enable more efficient architectures, as the algorithm would be able to exploit the full information content of the SNR time series rather than a down-sampled representation.

Future work will focus on extending EasyResNet to exploit deeper architectures and benchmarking against production-level LVK search pipelines to quantify gains in detection efficiency and computational cost.

## ACKNOWLEDGMENTS

This material is based upon work supported by NSF's LIGO Laboratory, which is a major facility fully funded by the National Science Foundation. The authors also gratefully acknowledge the support of the Science and Technology Facilities Council (STFC) of the United Kingdom, the Max-Planck-Society (MPS), and the State of Niedersachsen/Germany for support of the construction of Advanced LIGO and construction and operation of the GEO 600 detector. Additional support for Advanced LIGO was provided by the Australian Research Council. The authors gratefully acknowledge the Italian Istituto Nazionale di Fisica Nucleare (INFN), the French Centre National de la Recherche Scientifique (CNRS), and the Netherlands Organization for Scientific Research (NWO) for the construction and operation of the Virgo detector and the creation and support of the EGO consortium. The authors also gratefully acknowledge research support from these agencies as well as by the Council of Scientific and Industrial Research of India, the Department of Science and Technology, India, the Science & Engineering Research Board (SERB), India, the Ministry of Human Resource Development, India, the Spanish Agencia Estatal de Investigación (AEI), the Spanish Ministerio de Ciencia, Innovación y Universidades, the European Union NextGenerationEU/PRTR (PRTR-C17.I1), the ICSC—CentroNazionale di Ricerca in High Performance Computing, Big Data and Quantum Computing, funded by the European Union NextGenerationEU, the Comunitat Autonòma de les Illes Balears through the Conselleria d'Educació i Universitats, the Conselleria d'Innovació, Universitats, Ciència i Societat Digital de la Generalitat Valenciana and the CERCA Programme Generalitat de Catalunya, Spain, the Polish National Agency for Academic Exchange, the National Science Centre of Poland and the European Union—European Regional Development Fund; the Foundation for Polish Science (FNP), the Polish Ministry of Science and Higher Education, the Swiss National Science Foundation (SNSF), the Russian Science Foundation, the European Commission, the European Social Funds (ESF), the European Regional Development Funds (ERDF), the Royal Society, the Scottish Funding Council, the Scottish Universities Physics Alliance, the Hungarian Scientific Research Fund (OTKA), the French Lyon Institute of Origins (LIO), the Belgian Fonds de la Recherche Scientifique (FRS-FNRS), Actions de Recherche Concertées (ARC) and Fonds Wetenschappelijk Onderzoek—Vlaanderen (FWO), Belgium, the Paris Île-de-France Region, the National Research, Development and Innovation Office of Hungary (NKFIH), the National Research Foundation of Korea, the Natural Sciences and Engineering Research Council of Canada (NSERC), the Canadian Foundation for Innovation (CFI), the Brazilian Ministry of Science, Technology, and Innovations, the International Center for Theoretical Physics South American Institute for Fundamental Research (ICTP-SAIFR), the Research Grants Council of Hong Kong, the National Natural Science Foundation of China (NSFC), the Israel Science Foundation (ISF), the US-Israel Binational Science Fund (BSF), the Leverhulme Trust, the Research Corporation, the National Science and Technology Council (NSTC), Taiwan, the United States Department of Energy, and the Kavli Foundation. The authors gratefully acknowledge the support of the NSF, STFC, INFN, and CNRS for provision of computational resources. Part of our simulations utilized the Virtual Data cloud computing system at IJCLab. We thank Michel Jouvin and Gerard Marchal-Duval for their prompt support and advice about this system.

## DATA AVAILABILITY

The data that support the findings of this article are available upon request (due to significant size). All the codes are in this repo: [79].

## APPENDIX: CORRELATION BETWEEN THE TEMPLATES

The matched-filtering analysis consists of correlating the detector's output $s(t)$ with the templates $h(t)$, listed in the template bank, and weighting the correlation by the detector's noise power spectral density $S_n(f)$. The output of the matched filtering is the signal-to-noise-ratio time series $\rho(t)$, and this quantity shows a peak in correspondence with the merging time if the template and the signal are similar enough. The $\rho(t)$ of each template will encode the physical difference between the template and the signal present in the data. An efficient way to visualize it is to consider the spectrogram of a reference signal, with chirp mass $M_c^{\rm ref} = 2.17 M_\odot$ merging at $t_{\rm ref} = 40$ s and see how templates with chirp mass below and above act differently when compared with the reference one. For instance, suppose that a template with $M_c^1 = 0.95 M_\odot$ merges at $t_1 = 38$ s, and then the same template merges at $t_2 = 42$ s. In Fig. 15(a) we see that the template will cross the signal if it merges after the reference time. Conversely [Fig. 15(b)], considering instead a template with $M_c^2 = 4.35 M_\odot$, this time the behavior will be the opposite, and the template will cross the reference if it merges before $t_{\rm ref}$. This phenomenological picture describes qualitatively why we expect structures such as those shown in Fig. 1.

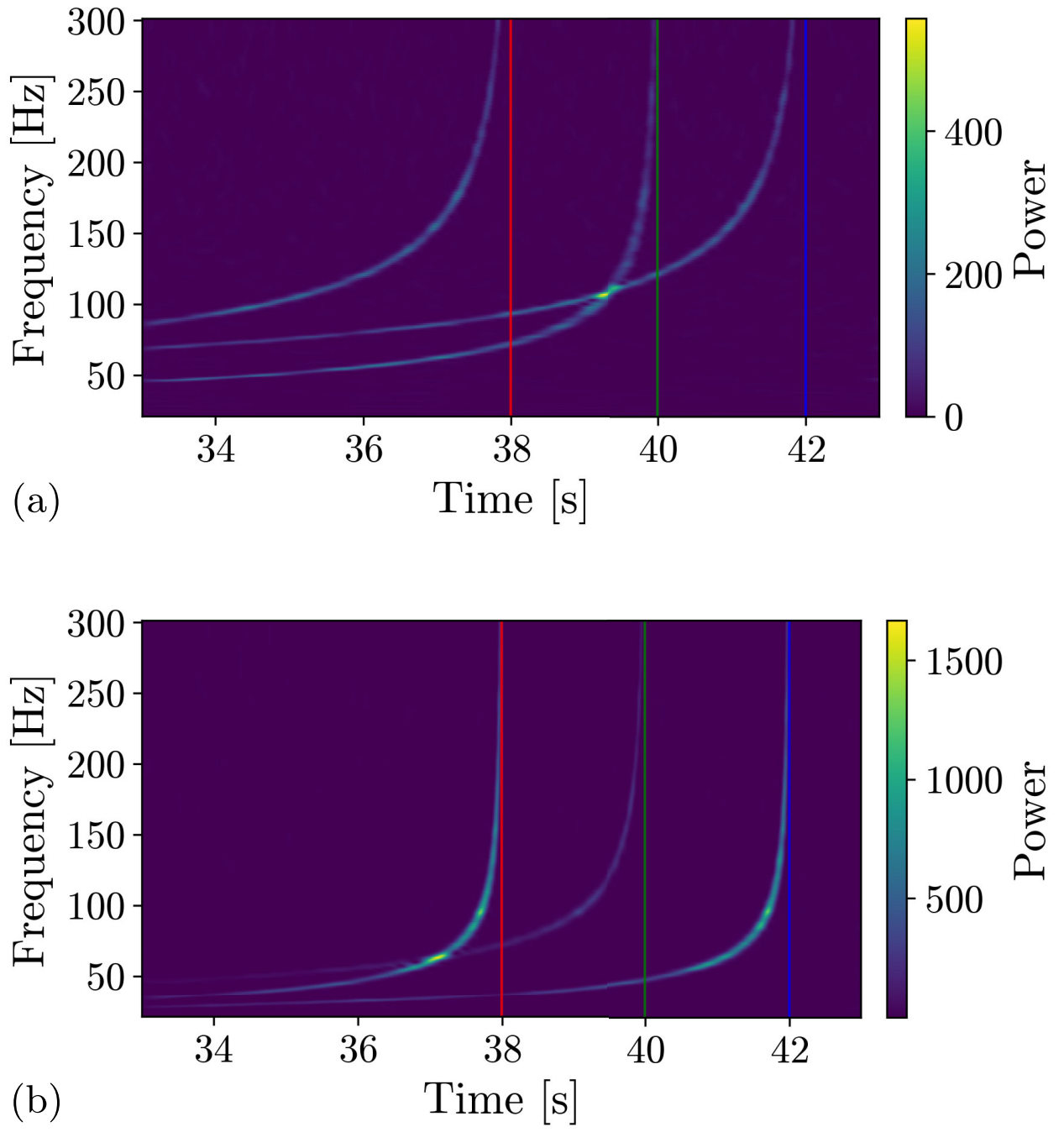

FIG. 15. Spectrograms comparing templates with chirp mass below (a) and above (b) the reference value $M_c^{\rm ref} = 2.17 M_\odot$, merging at $t_{\rm ref} = 40$ s (green line). (a) Template with $M_c^1 = 0.95 M_\odot$ merging at $t_1 = 38$ s (red line) and at $t_2 = 42$ s (blue line). (b) Same for a template with $M_c^2 = 4.35 M_\odot$. A template lighter than the reference crosses the signal track when it merges after $t_{\rm ref}$, whereas a heavier template crosses it when it merges before $t_{\rm ref}$. This qualitative behavior underlies the structures imprinted on the TT-SNR maps of Fig. 1.